\documentclass[sigconf]{acmart}

\AtBeginDocument{%
  \providecommand\BibTeX{{%
    \normalfont B\kern-0.5em{\scshape i\kern-0.25em b}\kern-0.8em\TeX}}}

\copyrightyear{2026}
\acmYear{2026}
\setcopyright{cc}
\setcctype{by}
\acmConference[CHI '26]{Proceedings of the 2026 CHI Conference on Human Factors in Computing Systems}{April 13--17, 2026}{Barcelona, Spain}
\acmBooktitle{Proceedings of the 2026 CHI Conference on Human Factors in Computing Systems (CHI '26), April 13--17, 2026, Barcelona, Spain}
\acmDOI{10.1145/3772318.3791689}
\acmISBN{979-8-4007-2278-3/2026/04}

\usepackage{graphicx}  
\usepackage{url}
\usepackage{booktabs}  
\usepackage{caption}
\usepackage{subcaption}
\usepackage{xspace}
\usepackage{multirow}

\graphicspath{{figures/}{pictures/}{images/}{./}}

\newcommand{\pp}{\textsc{PolicyPad}\xspace}

\begin{document}

\title[\pp for LLM Policy Prototyping]{\pp: Collaborative Prototyping of LLM Policies}

\author{K. J. Kevin Feng}
\affiliation{%
  \institution{University of Washington}
  \city{Seattle}
  \state{WA}
  \country{USA}}
\email{kjfeng@uw.edu}

\author{Tzu-Sheng Kuo}
\affiliation{%
  \institution{Carnegie Mellon University}
  \city{Pittsburgh}
  \state{PA}
  \country{USA}}
\email{tzushenk@cs.cmu.edu}

\author{Quan Ze (Jim) Chen}
\affiliation{%
  \institution{AI \& Democracy Foundation}
  \city{Seattle}
  \state{WA}
  \country{USA}}
\email{jim@aidemocracyfoundation.org}

\author{Inyoung Cheong}
\affiliation{%
  \institution{Princeton University}
  \city{Princeton}
  \state{NJ}
  \country{USA}}
\email{iycheong@princeton.edu}

\author{Kenneth Holstein}
\affiliation{%
  \institution{Carnegie Mellon University}
  \city{Pittsburgh}
  \state{PA}
  \country{USA}}
\email{kjholste@cs.cmu.edu}

\author{Amy X. Zhang}
\affiliation{%
  \institution{University of Washington}
  \city{Seattle}
  \state{WA}
  \country{USA}}
\email{axz@cs.uw.edu}

\renewcommand{\shortauthors}{Feng, et al.}

\begin{abstract}
  As LLMs gain adoption in high-stakes domains like mental health, domain experts are increasingly consulted to provide input into policies governing their behavior. From an observation of 19 policymaking workshops with 9 experts over 15 weeks, we identified opportunities to better support rapid experimentation, feedback, and iteration for collaborative policy design processes. We present \pp, an interactive system that facilitates the emerging practice of \textit{LLM policy prototyping} by drawing from established UX prototyping practices, including heuristic evaluation and storyboarding. Using \pp, policy designers can collaborate on drafting a policy in real time while independently testing policy-informed model behavior with usage scenarios. We evaluate \pp through workshops with 8 groups of 22 domain experts in mental health and law, finding that \pp enhanced collaborative dynamics during policy design, enabled tight feedback loops, and led to novel policy contributions. Overall, our work paves expert-informed paths for advancing AI alignment and safety.
\end{abstract}



\begin{teaserfigure}
  \centering
  \includegraphics[width=1\textwidth]{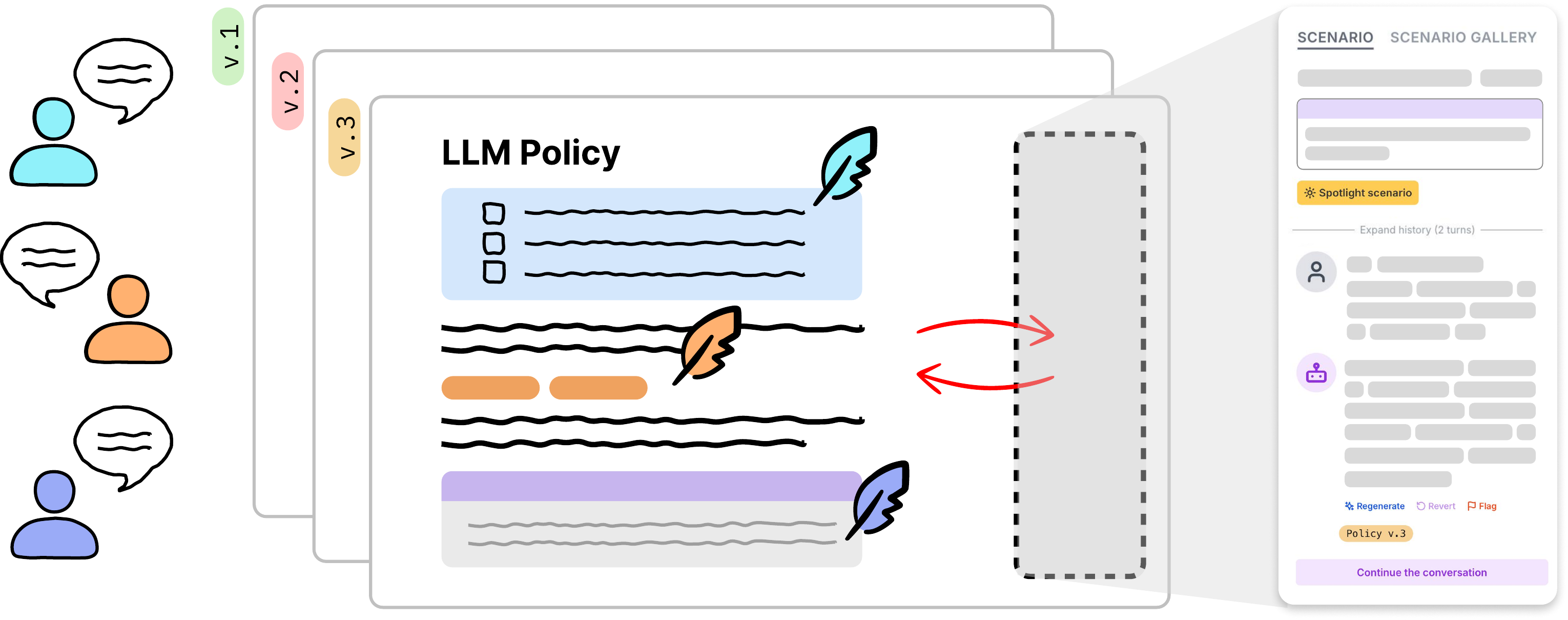}
  \caption{\pp is an interactive system that facilitates collaborative prototyping of LLM policies. Policy designers work together in real time \textbf{(left)} to draft policy statements in \pp's collaborative editor \textbf{(middle)}, while experimenting with the model's policy-informed behavior in a private sidebar \textbf{(right)}. Content from the private sidebar can be fluidly brought into the collaborative editor for viewing, editing, and discussion. To facilitate LLM policy prototyping, \pp borrows concepts and practices from UX prototyping, including heuristic evaluation, storyboarding, and rapid iteration.}
  \label{fig:teaser-abstract}
  \Description{Policy designers work together in real time (left) to draft policy statements in PolicyPad's collaborative editor (middle), while experimenting with the model's policy-informed behavior in a private sidebar (right). Content from the private sidebar can be fluidly brought into the collaborative editor for viewing, editing, and discussion. To facilitate LLM policy prototyping, PolicyPad borrows concepts and practices from UX prototyping, including heuristic evaluation, storyboarding, and rapid iteration.}
\end{teaserfigure}

\maketitle

\section{Introduction}

For decades, researchers and science fiction writers have imagined a world in which AI systems can be governed by natural language rules \cite{asimov1940robot, turing1950mind}. Today, governing large language models (LLMs) with \textbf{LLM policies\footnote{In this paper, our use of the term ``policy'' will refer to LLM policy (as opposed to public policies, government regulations, etc.) unless stated otherwise.}---sets of rules, guidelines, and desiderata that shape model behavior}---is a key component in the broader toolkit of approaches to improve model alignment and safety \cite{openai-collective-alignment, oai-model-spec, claude-constitution, huang2024collective, horwitz-meta-policy, lam2024ai}. For example, OpenAI's Model Spec contains a series of general objectives and principles (e.g., \textit{``Seek the truth together''}) for researchers and red-teamers to use as a guide when working on reinforcement learning from human feedback (RLHF) \cite{oai-model-spec}, as well as for the model to learn from directly \cite{guan2024deliberative}. Similarly, Anthropic uses Constitutional AI---in which reinforcement learning receives reward signals from AI-generated feedback that adheres to a set of principles (a ``constitution'')---to align its Claude models \cite{bai2022constitutional, claude-constitution}. If effective, LLM policies promise a transparent, familiar, and legible means by which developers and policymakers can govern AI systems \cite{he2025statutory, zhang2025stress, huang2024collective}.

As LLMs are deployed to millions of users globally, LLM policies become increasingly consequential and scrutinized. This is especially true in high-stakes, tightly regulated domains that are seeing rapid increases in LLM use by everyday users, such as mental health and law \cite{nyt-chatgpt-suicide, nyt-ai-lawyers, nyt-ai-therapists, nyt-psychosis-transcript, nyt-chatgpt-psychosis, cheong2024not, techcrunch}. LLM policies are primarily written by model developers, but the lack of expert input often leads to insular policies that risk delivering irresponsible model outputs to users in critical scenarios, while ignoring key safety concerns \cite{horwitz-meta-policy, nyt-psychosis-transcript, cheong2024not}.
While frontier model developers regularly partner with external domain experts to conduct pre-release safety testing of their models \cite{claude-4-syscard, hurst2024gpt, gemini-2.5-card}, there has been little documentation of similar efforts for LLM policies. Yet, there is mounting recognition that co-designing AI behaviors with experts is essential for LLM policy design, especially in safety-critical, domain-specific use cases \cite{oai-expert-input, Walsh2025VirologistOA}.

In this work, we first conduct a 15-week observational study in partnership with OpenAI to better understand how LLM policies can be co-designed with experts. Through 19 interactive workshops in which mental health experts discussed, annotated, taxonomized, and drafted user queries and LLM responses, we observed experts collaboratively ideating and discussing policy ideas while seeking ways to rapidly test and iterate on them through experimentation with model behavior on realistic scenarios. Much like prototyping in user experience (UX) practice, there is a strong emphasis on collaborative and rapid exploration, feedback collection, and iteration. We thus conceptualize this emerging practice as \textit{LLM policy prototyping}, borrowing from established UX practices like heuristic evaluation and low-fidelity prototyping. LLM policy prototyping draws upon and shares many motivations with LLM red-teaming and participatory model evaluation, but is specifically oriented towards producing an artifact that can subsequently inform red-teaming, model evaluation, and other safety efforts.

However, few tools exist for policy design \cite{lam2024ai}, let alone tools that support collaborative LLM policy prototyping. We observe notable opportunities for experts to tighten the feedback loop during policy design while leveraging collaborative affordances to build off each other's expertise. We design and develop \pp,\footnote{We open-source \pp at \url{https://github.com/kjfeng/policypad}.} an interactive system that facilitates LLM policy prototyping. \pp draws upon established methods and concepts within UX prototyping to enable small groups to collaboratively draft policies, test policy-informed model behavior against usage scenarios, evaluate the quality of the policy, and iterate on its contents in real time through tight feedback loops. 

We evaluate \pp through policy prototyping sessions with 22 domain experts spanning two domains---mental health and law---organized into 8 groups. We found that design decisions in \pp fostered collaborative dynamics between experts via its interactive in-editor widgets and yielded more novel policies compared to a baseline, relative to existing policies including OpenAI's Model Spec \cite{oai-model-spec} and Claude's Constitution \cite{claude-constitution}. Key areas of novelty include offering more specific guidelines on when the model should defer to a human expert, and eliciting key information required for responsible assistance early in the conversation. We end by discussing the practical implications of our work, including where LLM policy prototyping can be situated in AI alignment pipelines, and approaches for scaling up policy prototyping efforts.

Concretely, this work makes the following contributions:
\begin{itemize}
    \item A 15-week observational study with 9 mental health experts that surfaced opportunities for tight, collaborative feedback loops in LLM policy design.
    \item LLM policy prototyping, a conceptualization of an emerging practice for collaboratively prototyping LLM policies in small groups.
    \item \pp, a system that facilitates LLM policy prototyping through interactive and collaborative affordances for policy design, drawing from established UX practices.
    \item An evaluation of \pp with 22 domain experts in 2 domains, where we found that the system enriched collaboration during policy prototyping and resulted in more novel policies compared to a baseline.

\end{itemize}

\begin{figure*}
    \centering
    \includegraphics[width=0.9\linewidth]{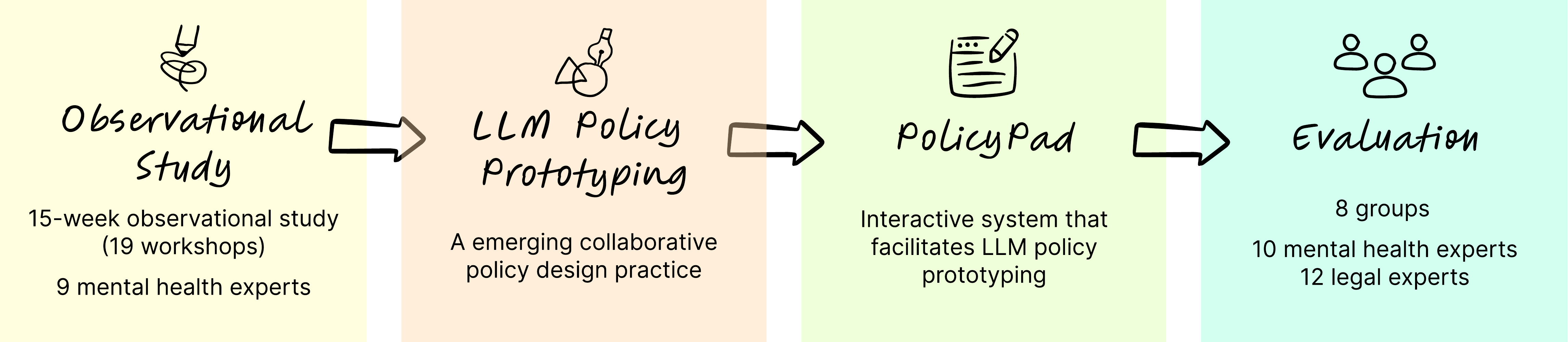}

    \caption{\textbf{Research Process Overview}. Our work proceeded in 4 phases: (1) a 15-week observational study with 9 mental health experts (19 workshops) led to (2) conceptualization of LLM policy prototyping. We then (3) designed and built \pp and (4) evaluated it through 8 policy prototyping sessions with 22 experts (10 mental health, 12 legal).}
    \Description{Four sticky notes in a line with arrows connecting them from left to right. Each sticky note describes a phase of our work. Our work proceeded in 4 phases: (1) a 15-week observational study with 9 mental health experts (19 workshops) led to (2) conceptualization of LLM policy prototyping. We then (3) designed and built \pp and (4) evaluated it through 8 policy prototyping sessions with 22 experts (10 mental health, 12 legal)}
    \label{fig:process}
\end{figure*}

\section{Related Work}

\subsection{Co-Designing AI Systems}

As AI systems simultaneously broaden and deepen their impact across society, there is an increasingly clear and urgent need to seek input on AI development from beyond AI developers \cite{suresh2024participation, delgado2023participatory, meta-community-forums, feng2025sociotechnical, huang2024collective, openai-collective-alignment, cheong2024not, feng2024canvil}. \textit{Co-design} provides an appropriate set of methods and frameworks to draw upon for diverse participants to collaboratively shape AI systems. Sanders and Stappers describe co-design as \textit{``the creativity of designers and people not trained in design working together in the design development process''} \cite{sanders2008co}. 

Co-design can be used for AI development and evaluation in many ways. For example, Lin et al. \cite{lin2021engaging} hosted co-design workshops with K--12 teachers to identify promising opportunities to integrate AI into core curricula. Frontier model developers have co-designed safety benchmarks for chemical and biological risks with virologists and national security experts through special partnerships with organizations like Gryphon Scientific \cite{claude-4-syscard, hurst2024gpt}. Overall, the `co' in `co-design' can refer to a variety of terms that capture the spirit of this approach, including collaborative, cooperative, collective, or connective \cite{zamenopoulos2018co}. In our work, we emphasize the \textit{collaborative} nature of co-design by supporting collaboration between domain experts and HCI researchers, as well as collaboration amongst domain experts themselves. 

Co-design is often discussed interchangeably with the related practice of participatory design (PD) \cite{zamenopoulos2018co, yu2025participatory}. While the two share similar aspirations and commitments, they have noteworthy differences in scope and procedure. Yu et al. \cite{yu2025participatory} argue that co-design encompasses a broader set of methods shaped by contributions from domains like product design, service design, and healthcare, while PD retains a more political identity, with strong commitments to democratic legitimacy, power redistribution, and long-term relational engagement. Indeed, these commitments are reflected in work in Participatory AI \cite{kuo2024policycraft, suresh2024participation, delgado2023participatory, meta-community-forums, barnett2025envisioning, ajmani2025secondary, sloane2022participation}. Additionally, participatory design creates a hybrid space that is neither in the stakeholders' nor facilitators' domain to allow collaboration to happen on ``even ground'' for all \cite{delgado2023participatory}, whereas co-design aims to explicitly draw upon the expertise of collaborating stakeholders \cite{long2021co, madaio2020co, lin2021engaging, tseng2025ownership}. Finally, co-design often focuses on designing products or artifacts \cite{kleinsmann2008barriers}, while participatory design may have a broader focus on sociotechnical systems that may contain those products or artifacts \cite{delgado2023participatory}.

Drawing from co-design, we work closely with domain experts in mental health and law to prototype LLM behavioral policies for those domains.
We contribute to existing literature on co-designing AI systems by focusing co-design efforts around an artifact that is increasingly recognized as central to AI alignment and governance strategies \cite{oai-model-spec, claude-constitution, he2025statutory, zhang2025stress}: LLM behavioral policies.

\subsection{Scenarios as Decision-Making and Anticipatory Tools}
Scholars across disciplines have found scenarios to be a valuable tool in their research and practice. Two complementary uses of scenarios are to 1) directly support decision making, and 2) help anticipate potential future outcomes that can then lead to more informed decisions. 

Scenarios have been shown to be helpful in grounding human decision-making, \textit{case studies} are scenario-based curricula commonly used to train practitioners to sharpen their decision-making abilities in fields such as law, medicine, and business \cite{molewijk2008teaching, farashahi2018effectiveness, feng2023case}. In the legal domain specifically, cases represent past court decisions and are foundational for \textit{case law}, in which courts rely on rulings in precedent cases to resolve new ones \cite{caputo2024alignment, kagan2019adversarial}. In collaborative settings with more than one decision-maker, cases provide contextual details and nuances that serve as a medium for productive deliberation to establish common ground and reconcile differences \cite{ontanon2006arguments, glez2002analytical}. For example, PolicyCraft \cite{kuo2024policycraft} is an interactive system that uses case-based deliberation and voting to scaffold participatory policy design. Other applications of case-based deliberation include legal adjudication \cite{carpenter1917court}, medical diagnosis \cite{schaekermann2019understanding}, and content moderation \cite{fan2020digital, pan2022comparing}.

Scenarios are also useful for envisioning potential uses of technology and anticipating technology's societal consequences. Scenario-based design and storyboarding are commonly used early in the human-centered design process as provocations for designers envision ways a technology can meet user needs \cite{figma-storyboards, carrol1999five, bodker1999scenarios, hooper1982scenario, kuo2023understanding}. Barnett et al. \cite{Barnett2025Scenarios} surveyed scenario-building methods in computer science research over the past decade and identified five main ways computing researchers use scenarios: 1) to gather stakeholder needs and values, 2) to empower marginalized groups to imagine
technology futures, 3) to provoke ethical reflection and promote critical awareness, 4) to anticipate threats and risks of these technologies, and 5) to explore perceptions and
impacts of novel technologies before they launch. Interactive systems have been developed by the HCI community (e.g., Blip \cite{pang2024blip}, Farsight \cite{wang2024farsight}) to help developers anticipate impacts of their work by retrieving existing scenarios or generating new ones.  

In our work, we primarily use scenarios to aid expert deliberation and decision-making on what appropriate model behavior should look like in high-stakes, domain-specific contexts. We develop our scenarios by sampling datasets of realistic cases in mental health and legal practice introduced by prior work \cite{lamparth2025moving, cheong2024not}. Our work builds on prior work by enabling \textit{interactive scenarios} that experts can extend, collaboratively analyze, and use to interrogate model behavior. As they do so and work with others to collaboratively design an LLM policy, experts may also become better equipped to anticipate consequences and risks of LLM use in their domains.

\subsection{Human-Centered Approaches to AI Alignment and Safety}
\label{s:rw-redteaming-policy}
AI alignment broadly refers to efforts to align the behavior of AI systems to human preferences and values \cite{gabriel2020artificial, ngo2022alignment}. Relatedly, AI safety aims to reduce risks that arise from the development and deployment of AI systems \cite{kasirzadeh2024two, anderljung2023frontier, bengio2025international}. Researchers increasingly recognize the importance of human-centered approaches, grounded in users' personal and collective needs, in both areas \cite{terry2024interactive, deng2025weaudit, morris2025hci, feng2024canvil, lam2024ai, feng2025sociotechnical, kuo2023understanding}.

A growing body of work responding to this recognition focuses on empowering non-AI experts to surface problematic AI behavior through evaluating, auditing, and red-teaming \textit{model outputs} \cite{deng2025weaudit, lam2022end, devrio2022toward, shen2021everyday}. For example, DeVrio et al. \cite{devrio2022toward} investigated everyday users' strategies for uncovering harmful algorithmic behavior and found that 1) these strategies were strongly influenced by personal experiences with societal bias, and 2) collaborative sensemaking between multiple users is a promising approach for user-driven audits. WeAudit \cite{deng2025weaudit} presents a workflow and platform for everyday users to collectively audit text-to-image models and generate an audit report with actionable insights to AI developers.
Developers have also employed hybrid human-AI red-teaming campaigns \cite{claude-4-syscard, gemini-2.5-card, hurst2024gpt, zhang2025work} or purely automated ones \cite{perez2022red, zhou2025autoredteamer}. Across all of these approaches, prompt engineering has been extensively used as a technique to probe model behavior \cite{kim2024evallm, arawjo2024chainforge, ganguli2022redteaminglanguagemodels}.

Auditing and red-teaming at the output level is often a game of whack-a-mole: there is seemingly an endless stream of problematic behaviors to patch and new ones are constantly surfacing. Researchers have thus developed complementary approaches where they define \textit{higher-level behavioral policies}---often called ``model specs'' \cite{oai-model-spec} or ``constitutions'' \cite{claude-constitution}---and train them into the model \cite{bai2022constitutional, huang2024collective, openai-collective-alignment, guan2024deliberative}. These policies can be a promising artifact model behavior governance because they 1) can be easily authored and understood by people \cite{he2025statutory, zhang2025stress}, 2) can serve as guidelines for human red-teamers and data annotators to improve other parts of the alignment pipeline \cite{oai-model-spec}, and 3) are shown to be effective in inducing more safe and aligned model behavior, even \textit{before} the model is red-teamed \cite{guan2024deliberative, bai2022constitutional}. However, these policies are not without the limitations. Vague or poorly written policies can be misinterpreted by the model \cite{he2025statutory}, or policies may contain conflicting statements that give the model noisy signals during training \cite{zhang2025stress}. As such, designing these policies should be an iterative and continual process \cite{openai-collective-alignment}.

Our work focuses on iteratively improving the design of LLM behavioral policies by inviting experts to contribute to them directly. While our primary focus is policy design, we are also inspired by approaches taken by literature on end-user model auditing and red-teaming. Specifically, we design affordances for experts to collaboratively test, critique, and discuss model outputs in response to policy changes to help them envision and implement policy improvements.

\subsection{Tools for LLM Policy Design}
Due to the nascency of LLM policy design, there are currently few tools to support policy designers, despite the rising importance of LLM policies \cite{lam2024ai, huang2024collective}. Policy Projector by Lam et al. \cite{lam2024ai} supports AI safety practitioners in authoring if-then rules for LLM content moderation. Our work differs in that we specifically support collaborative policy design, which Lam et al. identified as a fruitful area of future work. ConstitutionMaker \cite{petridis2024constitutionmaker} allows users to turn written critiques of model responses into principles that guide future behavior, but is a tool for LLM personalization rather than policy design. Roleplay-doh \cite{louie2024roleplay} similarly converts written feedback on LLM behavior into principles, but specifically for domain experts to govern LLM-prompted roleplay. We see this feedback-to-principle interaction as valuable to policy designers as well, and integrate a version of it into our system.

\section{Observational Study}
\label{s:formative}

\subsection{Study Motivation and Procedure}
To develop an understanding of real-world LLM policy design practices, we conducted a 15-week-long observational study via contextual inquiry \cite{beyer1999contextual} in partnership with OpenAI. We wanted to observe how domain experts collaborated to draft domain-specific LLM policies,\footnote{Recall that an LLM policy is a set of rules, guidelines, and desiderata that shape model behavior.} and any opportunities for improving processes and/or tooling. 

We acknowledge that ``expertise'' is a contested topic, especially in democratic decision-making contexts \cite{diaz2024expertise}. We use a narrow definition of expert and recruit participants who have 1) at least two years of practical experience conducting client-facing work in their domain, and 2) are pursuing or have completed an advanced degree (Master's or PhD or equivalent). This combination ensures that policy design is guided by both participants' real-world experiences and aspects of their formal training. 

\subsubsection{Background and participants}
OpenAI, in collaboration with a small group of external researchers including us, was organizing weekly/twice-a-week virtual workshops (19 workshops total) with 9 experts in clinical mental health (denoted E1--E9) to design new AI policies for model behavior when responding to users' mental health queries. While all 9 experts were invited to every workshop, there were not 9 attendees every week due to scheduling conflicts. Out of the experts, 6 identified as female and 3 as male. For their highest degrees, 4 held a Ph.D. in clinical psychology, 4 held a Doctor of Clinical Psychology (Psy.D.), and 1 held a Master's in Clinical Psychology. Experts were all based in the United States. Participant recruitment and payment were handled by OpenAI and its data partners. Data from the workshops were shared among all collaborators, but our data analysis procedures and research outputs were distinct from those of the company. This study was classified as exempt by the University of Washington IRB. 

\subsubsection{Procedure}
At least one member of our research team attended these workshops from January to April 2024. Each workshop was 60--90 minutes in length. One facilitator from either the AI lab or our research team led the workshop with a collaborative policy design activity for the group of experts. These activities revolved around two goals. First, experts developed taxonomies for collections of example mental health-related user queries to an LLM (``scenarios''). Some specific tasks for this goal included deliberating with other experts on taxonomy labels, when to combine and separate labels, and assigning labels to scenarios. Second, experts drafted and voted on desirable rules the model should follow. This included tasks such as reviewing proposed rules in a shared spreadsheet, suggesting modifications, and merging similar rules. The agenda for each workshop was set by either OpenAI researchers or one of its collaborators. Facilitators' guides for the workshops we organized can be found in Supplementary Materials.

\begin{figure*}[t]
    \centering
    \includegraphics[width=0.9\linewidth]{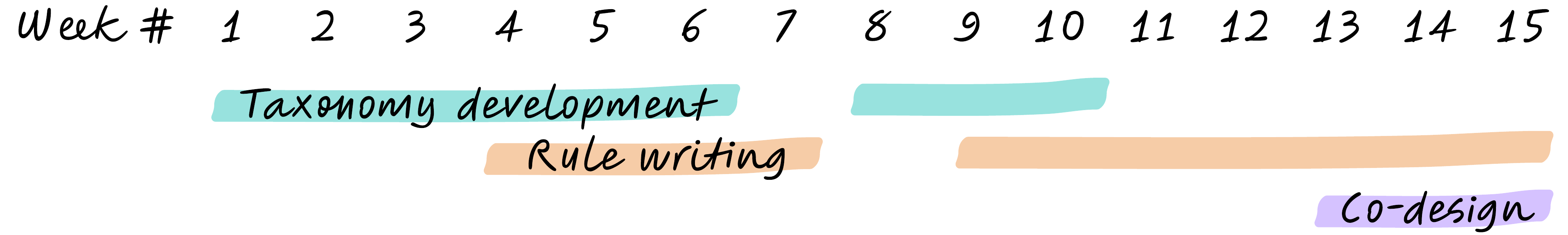}
    \caption{Timeline of activities in our 15-week observational study. During \textit{taxonomy development}, experts organized and taxonomized a collection of diverse LLM usage scenarios. During \textit{rule writing}, experts drafted, discussed, and refined rules to govern LLM behavior. During \textit{co-design}, experts interacted with and gave feedback on prototypes we built of a tool for collaborative policy design. There was fluid movement between taxonomy development and rule-writing, such that some sessions included activities pertaining to both goals.}
    \Description{Weeks represented by numbers 1 to 15 are listed out horizontally. Highlighting underneath shows that taxonomy development spanned weeks 1--6 and 8--10. Rule writing spanned weeks 4--7 and 9--15. Co-design spanned 13--15.}
    \label{fig:obs-timeline}
\end{figure*}

Experts generally tackled the first goal in earlier weeks and the second in later weeks. There was, however, fluid movement and substantial iteration across the two goals, such that some workshops contained activities pertaining to both goals. As the weeks passed, we also moved towards co-designing tooling\footnote{Note that this specifically refers to co-designing tooling for collaborative policy design. It is in service of our broader goal of co-designing policies with experts.} with experts as we developed a better understanding of the problem space and experts' pain points. As part of the final three workshops, we reserved some time to show experts initial prototypes of an interactive system to assist with collaborative LLM policy design, collected their feedback, and iterated on the prototype for the following week (see Appendix \ref{a:co-design} for prototype iterations). In the final workshop, we conducted a 30-minute semi-structured exit interview with all experts, asking them to reflect on the workshops and their policy contributions. Figure \ref{fig:obs-timeline} shows an overview of activities across the 15-week period. 

\subsubsection{Data analysis}
All workshops were recorded and transcribed. As is common in contextual inquiry, team members took observational notes and asked questions as needed \cite{beyer1999contextual}.
The first author then deductively coded workshop transcripts based on themes identified in our team's notes, clarifying and iterating on the themes while doing so. The first author used a hybrid inductive-deductive process \cite{fereday2006demonstrating} to code the transcript of the exit interview. This hybrid process allowed us to connect to themes from our workshop data while embracing new themes emerging from the interview. Our final set of themes can be found in Table \ref{t:formative-themes} in Appendix \ref{a:formative-themes}. 

\subsection{Observational Study Results}
\label{s:form-opportunities}

We used our notes and themes from all workshops, including our exit interview, to synthesize four main observations, which we describe in detail below. Henceforth, we use the term ``\textbf{policy-informed model}'' to refer to an LLM that has been instructed to act in accordance to the policy drafted by experts.

\subsubsection{Incomplete feedback loops without model experimentation}
\label{form:prototyping}
Throughout the workshops, experts had visions for how their contributions to the policy through drafting taxonomies and principles can impact model behavior. For example, E3 shared that \textit{``a clinical minimization [of the user's feelings] can be helpful, but for the model, that would be hard to decipher''} so the group wrote a policy barring the model from engaging in this behavior. However, they failed to verify whether and how those visions \textit{actually came into fruition} because they did not have a policy-informed model to interact with. This resulted in an incomplete feedback loop. Experts were unable to obtain signals about the effectiveness of their policy contributions and any unintended side effects that may arise, as E9 explains: \textit{``Just because you think that might be a good rule, it may have an unanticipated consequence you don't realize. I think that it would be really helpful to know how these [rules] we're coming up with actually play out.''} E7 agreed and added that direct experimentation with a policy-informed model would allow them to better \textit{``see how [a conversational interaction] would play out from the perspective of a user.''} Experts had unrestricted internet access throughout the workshops and could test behaviors out on popular chatbots. However, we did not observe instances of this, possibly due to preoccupation with workshop activities, or lack of knowledge about (or in some systems, inability to set) custom system prompts.

\subsubsection{Experts tackled both high-level strategy and low-level semantics}
\label{form:heuristics}
We noticed that experts could easily derail from workflows that would enable them to best contribute their expertise when designing policies. For example, when creating taxonomies for mental health-related user queries, experts spent substantial time wrestling with wording and semantics. 
Similarly, E9 reflected that much of their time was spent on finding the right wording for taxonomy labels: \textit{``we thought needed to not spend forever trying to wordsmith exactly how that needed to appear.''} In a separate activity where experts wrote out ideal model responses, E5 agreed that experts should avoid getting stuck in the weeds of low-level wording edits: \textit{``It would be more effective at this stage for us to just put our thoughts in about what's right or wrong, because the time it takes to craft the perfect response is out of scope for this task.''} While important, study facilitators agreed that much of the low-level semantics of the policy can be refined post-hoc via LLMs, as long as there are sufficient amounts of expert insight to guide that refinement.

\subsubsection{Scenarios grounded discussions and spurred policy generation}
\label{form:scenarios}
We found that experts engaged in richer discussions that led to insightful policy suggestions after they were given scenarios, or examples of user-AI conversations that may arise in real-world use, for reference. For E1, looking at scenarios helped them identify two pieces of information the model should consider in its response: \textit{``We need to ask clarifying questions, in particular to clarify the severity and the nature of the dark thoughts this person suggested. Another dimension is to identify how long they've been feeling this way and what sources of support they have.''} E2 agreed with the need for a severity assessment, suggesting a safety rating scale for the user in case they cannot quickly reach a professional and need an immediate response: \textit{``The AI needs to respond, providing resources quickly. Maybe having a rating scale on the scale of zero to 10, how safe are you feeling right now?''} Adding on, E3 suggested eliciting the user's financial ability to pay for therapy and making referrals accordingly: \textit{``There might be questions instead like, what is your financial ability to pay for therapy right now? And if it's within certain ranges, then you might make a community mental health referral, like here's some Medicare people in your area.''} Exploring scenarios helped experts spot recurring problems in model responses and turn them into clear policies. While rules should stay broad enough to be useful, it is unclear how specific they should be. When scenarios show patterns that keep causing issues, they become obvious candidates for new rules, as E7 describes: \textit{``I keep seeing this thing over and over and it's incorrect, so that needs to be a rule.''} 

\subsubsection{Experts valued synchronous collaboration}
\label{form:synchronous}
In contrast with prior work that collected human feedback via asynchronous annotation (e.g., \cite{ouyang2022training, bai2022constitutional}) and/or focused on asynchronous policy design \cite{lam2024ai, kuo2024policycraft}, our workshops engaged experts in \textit{synchronous} collaboration---drafting, discussing, and iterating on policy in real time. Experts unanimously agreed that synchronous collaboration was enjoyable and productive. In E1's words, \textit{``I found it hugely rewarding and beneficial personally and professionally [...] I think we can get stuck in our heads because we're working on our own with our clients so much. It was really nice to hear other people's perspectives and thoughts.''} E6 emphasized the support and learning opportunities afforded by collaboration: \textit{``[it was] very supportive having other voices in the back of your head [...] it's been incredible learning with everyone.''} E9 found synchronous collaboration important for surfacing new perspectives and broadening coverage of the policy: \textit{``[...] there were times where someone else said something that just never occurred to me. We all know one person's opinion is never sufficient, especially in an area as diverse as mental health.''} Broadly, we observed that experts were able to quickly resolve disagreements and draft policy statements that had broad consensus in a synchronous setting.

\section{LLM Policy Prototyping}
\label{s:policy-prototyping}
Our observations in Section \ref{s:formative} posed challenges that are not foreign to HCI; well-established concepts and methods in UX design and prototyping can offer help in mitigating these challenges. Indeed, scholars have been increasingly interested in applying prototyping principles to designing bills and other public policies \cite{kimbell2017prototyping, hagan2021prototyping, privacy-policy-prototyping, quicksey-policy-prototypes, Kontschieder-policy-prototyping}, but minimal efforts have been made to expand these applications to beyond policies for governments. Additionally, while literature in end-user model auditing and red-teaming has allowed non-AI experts to identify problematic model behaviors (see Section \ref{s:rw-redteaming-policy}), few methods exist for directing outcomes of efforts to meet increasing demand for iteratively improving LLM policy design \cite{he2025statutory, zhang2025stress, huang2024collective, openai-collective-alignment}. Our work seeks to fill this need through co-designing policies with experts. 

We now describe \textbf{LLM policy prototyping} (henceforth ``policy prototyping'' for brevity), an emerging practice by which groups of individuals can synchronously collaborate on designing an LLM behavioral policy. Specifically, we map observations we identified in Section \ref{s:form-opportunities} to relevant UX methods, which are then mapped to their usage in policy prototyping. For example, to address Section \ref{form:prototyping}, \textit{enabling tight feedback loops} between ideation, design, and testing allows policy designers to quickly identify ``usability'' issues like unclear policy statements, explore policy designs that more effectively achieves desired model behavior, and work collaboratively to address issues as they surface. For Section \ref{form:heuristics}, experts' efforts can be better focused by conducting \textit{heuristic evaluation} on a policy using heuristics that direct experts' attention to higher-level desiderata (e.g., does the policy draw from real-world practices in their field?) rather than low-level wording edits. Promising uses of scenarios in Section \ref{form:scenarios} can be further scaffolded with techniques inspired by \textit{storyboarding}, where each conversational turn in the scenario grounds policy design ideas in concrete representations of users, contexts, and tasks.
Our full mapping is depicted in Table \ref{t:mapping} in Appendix \ref{a:ux-polipro}. 

In this work, we focus on \textit{low-fidelity} policy prototypes---artifacts with the primary goal of eliciting and integrating group perspectives on responsible model behavior, rather than a high-fidelity, ``production-ready'' policy. We leave the translation of low- to high-fidelity policies to future work.

\begin{figure*}[h]
    \centering
    \includegraphics[width=0.8\linewidth]{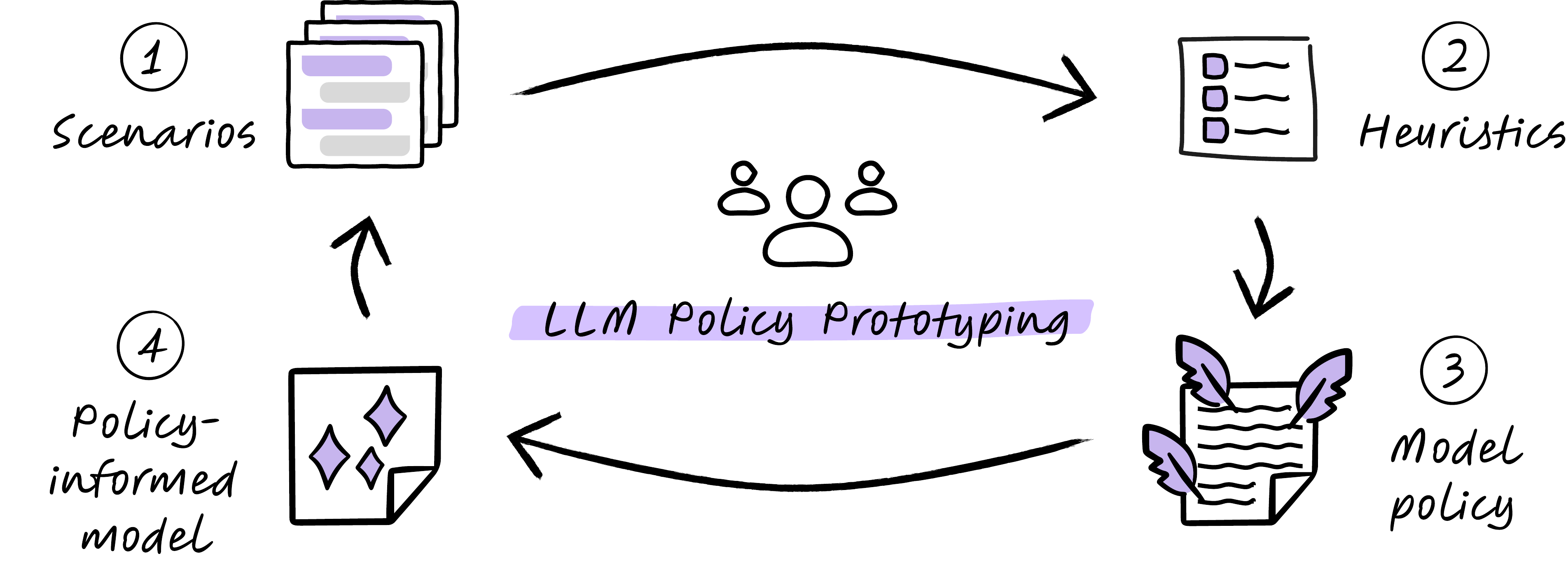}
    \caption{Illustration of our envisioned \textit{LLM policy prototyping} process. Scenarios inform desiderata for the policy via heuristics, which in turn guide the design of the policy. The policy shapes the behavior of a policy-informed LLM, which designers can then test against the scenarios to observe changes in behavior. The process is iterative: feedback from testing may lead to the creation of new scenarios, heuristics, and policy statements.}
    \Description{A figure illustrating the flow of LLM policy prototyping: Scenarios (1) inform desiderata for the policy via heuristics (2), which in turn guide the design of the policy (3). The policy shapes the behavior of a policy-informed LLM (4), which designers can then test against the scenarios to observe changes in behavior. The process is iterative: feedback from testing may lead to the creation of new scenarios, heuristics, and policy statements.}
    \label{fig:polipro-flow}
\end{figure*}

Concretely, we propose policy prototyping for a policy $P$ in domain $D$ to involve the following activities, as illustrated in Figure~\ref{fig:polipro-flow}.  
\begin{enumerate}
    \item Policy designers review a small set of scenarios that accurately depict real-world use of AI in a specific domain. This allows designers to better understand current AI behavior and contexts of use. 
    \item Informed by the scenarios, designers finalize a set of heuristics they will use to ensure $P$ preserves its quality and focus across many iterations.
    \item Guided by the heuristics, designers collaboratively draft policy statements that they believe will lead to more responsible model behavior in $D$. 
    \item Designers test a policy-informed model that acts in accordance with $P$ with existing scenarios. New scenarios, heuristics, and policy statements may be created based on insights from testing and discussions with other designers. The process then repeats until the prototyping session concludes. 
\end{enumerate}

\section{PolicyPad}
\label{s:system}

We introduce \pp, an interactive and collaborative system for synchronous LLM policy prototyping. We first describe how we arrived at our final design through three co-design sessions. Then, we walk through the system and its features.

\subsection{Iterative Co-Design Sessions with Experts}
\label{s:system:co-design-session}

We designed \pp iteratively through co-design sessions with the same participants in our observational study. These sessions were conducted towards the end of our observational study period. In each session, we presented an interactive prototype of the system.\footnote{Two prototypes were in Figma. One was implemented in TypeScript and React.} We then collected semi-structured feedback from participants and iterated on the prototype based on feedback for the next session. We repeated this until data saturation---participants were no longer able to provide substantial feedback until we implemented the system, for a total of three sessions (c.f. \cite{Kruzan2022DevelopingAMA}). 

Our \textbf{first prototype} consisted of a simple collaborative editor for policy authoring, along with a sidebar that allowed experts to engage in conversation with the policy-informed model. Experts' appreciation of the realtime collaborative features and desires for  more structured and systematic ways to interact with scenarios led us to the next iteration. Our \textbf{second prototype} introduced a more sophisticated side panel that allowed users to browse scenarios and use them to test model behavior. While experts agreed that this was a significant improvement, they desired closer integration between policy editing and scenario exploration. In our \textbf{third prototype}, we integrated scenarios into the editor as interactive widgets, while keeping the sidebar as a place to view scenario-specific information. Experts appreciated this and suggested ways to flag problematic model responses from the sidebar, as well as cleaning up unnecessary elements from the sidebar. We incorporated these suggestions into our final system. Detailed documentation on how we integrated participants' feedback across versions, along with prototype screenshots, can be found in Appendix \ref{a:co-design}. In the following section, we illustrate \pp's capabilities using a system walkthrough.

\subsection{System Walkthrough}
\pp can be used by any individual or group who wishes to facilitate a policy prototyping session---whether it be an AI lab, academic group, non-profit, or another organization. To do so, we use a running example of hypothetical LLM policy for financial advice. As a note on this section's terminology, we distinguish ``facilitators'' (those running the policy prototyping session) from ``users'' (policy designers participating in the session).

\begin{figure*}[t]
    \centering
    \includegraphics[width=1\linewidth]{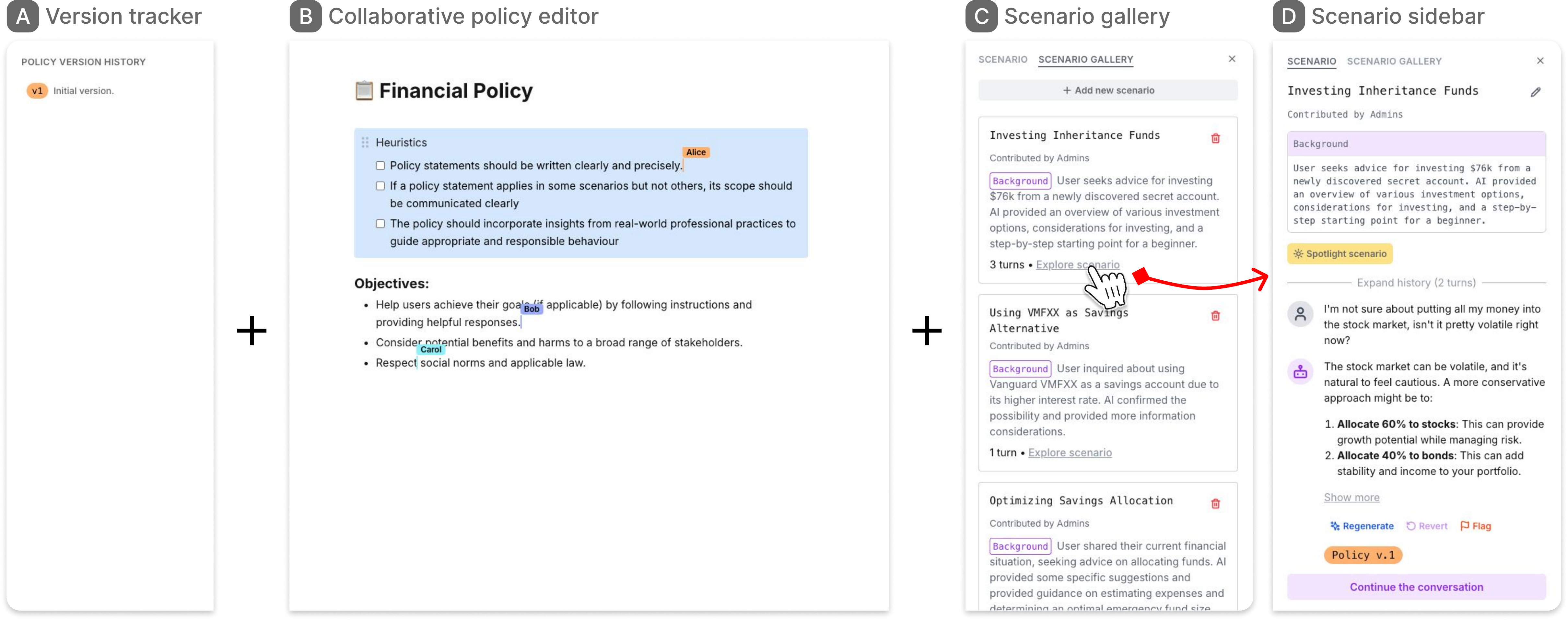}
    \caption{Main components of the \pp system. Users can keep track of their policy version in the left sidebar \textbf{(A)} as they collaborative edit the policy in the editor \textbf{(B)}. Users can access scenarios via the scenario gallery \textbf{(C)}. When they click into a scenario, they can view its full details and explore how the policy-informed model will behave on it using the scenario sidebar \textbf{(D)}.}
    \Description{Main components of the \pp system. Users can keep track of their policy version in the left sidebar (A) as they collaborative edit the policy in the editor (B). Users can access scenarios via the scenario gallery (C). When they click into a scenario, they can view its full details and explore how the policy-informed model will behave on it using the scenario sidebar (D).}
    \label{fig:pp-components}
\end{figure*}

\subsubsection{Preliminaries}
When users log into \pp, they see a collaborative document editor (Fig \ref{fig:pp-components} \textbf{B}), similar to Google Docs. The facilitator may provide light starting materials for the policy, such as high-level objectives, an initial set of policy heuristics (perhaps drawn from trust \& safety literature), and a few scenarios for the group to work with. Ideally, scenarios are representative of real-world model use in a domain. For example, facilitators who have access to chatbot logs may use privacy-preserving conversations from their logs. 

Users can access these scenarios via the \textbf{scenario gallery} in the right sidebar (Fig. \ref{fig:pp-components} \textbf{C}). While the document is a collaborative workspace, the sidebar is private to each user, allowing for independent experimentation with the policy-informed model.

\subsubsection{Scenario sidebar}
A user can browse the scenarios in the gallery and open a scenario in a detailed view (Fig. \ref{fig:pp-components} \textbf{D}). The scenario expands to fill the sidebar with the full user-AI conversation, as well as a brief, AI-generated summary\footnote{Generated with GPT-4o. More technical details are in Section \ref{s:technical}.} of the conversation's contents thus far. As the group reads the scenarios, they start to develop ideas for what to include in the policy.

Formally, a scenario in \pp comprises three parts: the background (all messages in the conversation up until the most recent turn), the newest user message, and the newest AI message. Given the background and the newest user message, the policy-informed model\footnote{The policy-informed model is an instance of Llama 3.3 70B Instruct. More technical details are in Section \ref{s:technical}.} generates the newest AI message. 

\subsubsection{Interactive in-editor scenario widgets}

\begin{figure*}[h]
    \centering
    \includegraphics[width=1\linewidth]{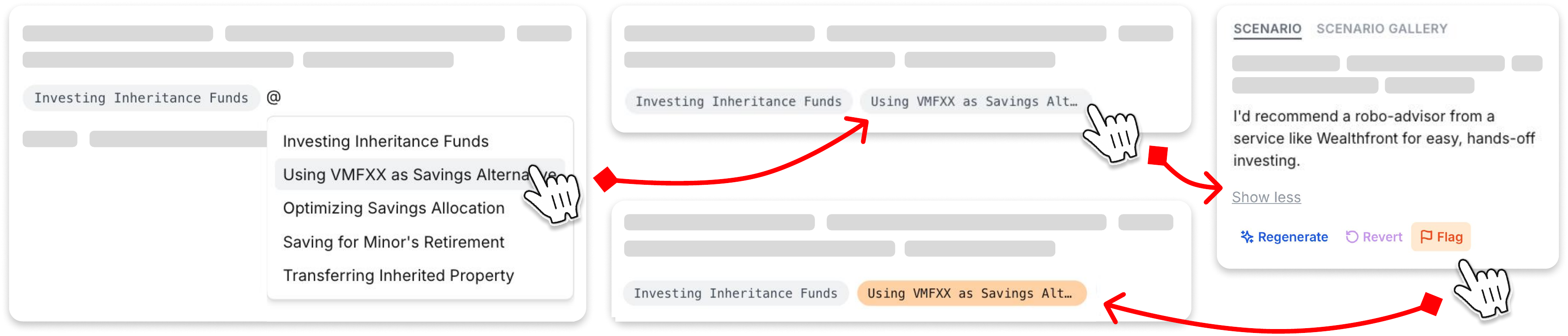}
    \caption{Scenarios can be brought into the editor inline with the policy as interactive widgets via referencing the scenario's title with the '@' symbol. Once in the editor, all users can click on it, view it in their scenario sidebar, and flag responses for group discussion.}
    \Description{Scenarios can be brought into the editor inline with the policy as interactive widgets via referencing the scenario's title with the '@' symbol. Once selected, it appears inline in the editor alongside text and other widgets. Once in the editor, all users can click on it, view it in their scenario sidebar, and flag responses for group discussion. If flagged, the widget will glow orange for everyone.}
    \label{fig:scenario-widgets}
\end{figure*}

Users can bring a scenario from their own scenario sidebar into the collaborative editor by referencing its title with the '@' symbol. Once referenced, the scenario appears inline in the editor as an interactive, pill-shaped widget (Fig. \ref{fig:scenario-widgets}). When a user clicks the widget, the full scenario will be shown in their scenario sidebar. These widgets can be used as illustrative examples of model behavior and build shared context when designing the policy. For example, a user may observe that the model does not provide disclosures of capability limits, or incorrectly assumes a detail not explicit in the conversational context. They can flag a model response in their scenario sidebar, which will make the scenario widget glow orange in the editor, encouraging others to take a look. 

\subsubsection{Drafting policy statements}

Once a group reviews the heuristics at the top of the editor to ensure they have a common understanding of desired policy goals, they are ready to start drafting policy statements. These policies address oddities, concerns, and other noteworthy aspects of model behavior they observed and flagged in the scenarios.

As an example, for a policy on providing responsible financial advice, a user may add a policy statement instructing the model to use \textit{cautious, neutral, and non-prescriptive language} while \textit{always surfacing a brief disclosure of limitations early in the conversation}. A couple users may collaboratively draft a policy statement for the model to \textit{defer the user to a licensed adviser or a compliant robo-advice product that meets regulatory obligations}. The group can review the policy together and take advantage of the real-time collaborative editing features to further refine each other's statements.

\subsubsection{Testing the policy with scenarios}
\begin{figure*}[h]
    \centering
    \includegraphics[width=0.7\linewidth]{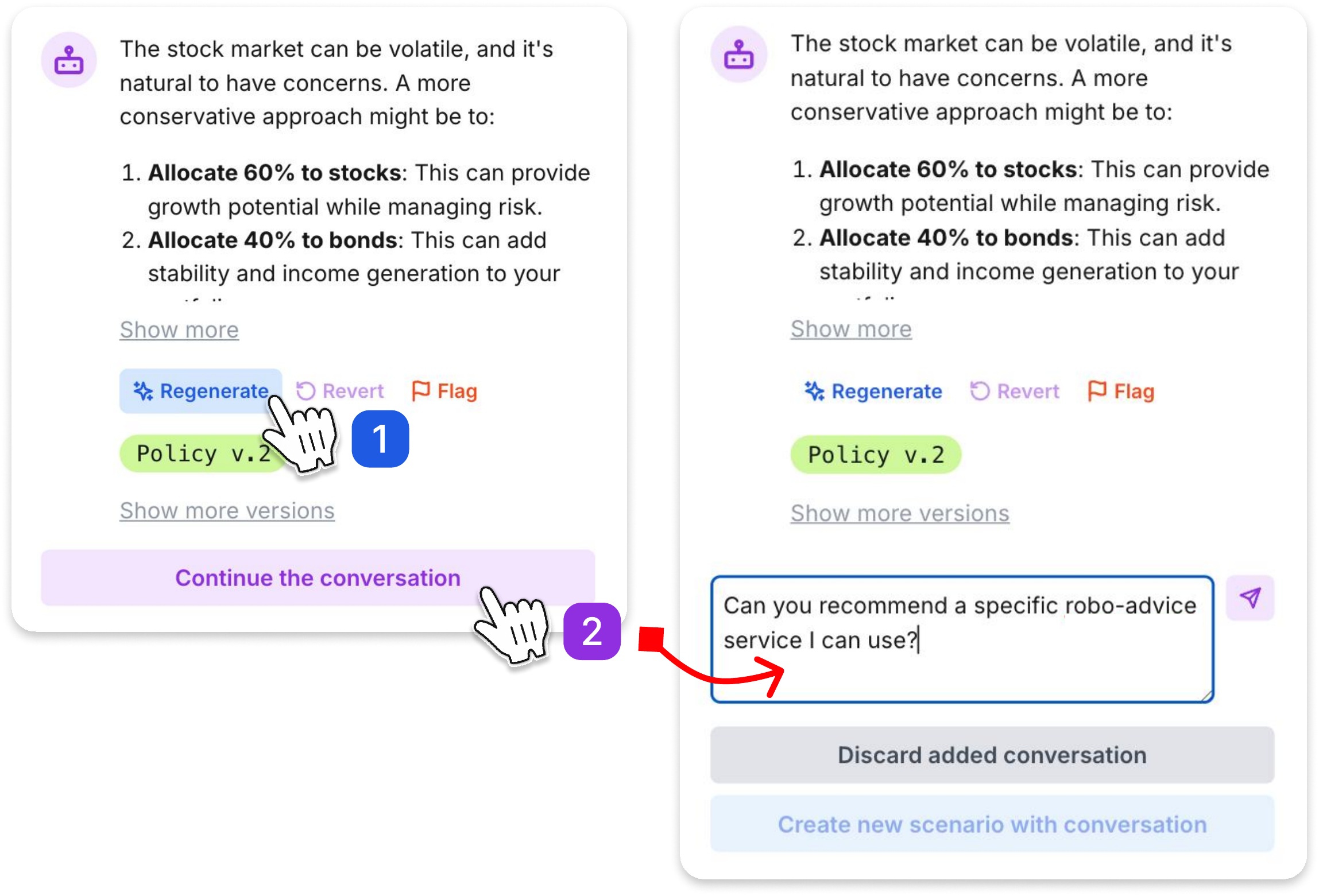}
    \caption{\pp offers two ways to test the behavior of the policy-informed model against a scenario: \textbf{(1)} regenerating the latest AI message, or \textbf{(2)} continuing the conversation.}
    \label{fig:testing-policy}
    \Description{UI showing two ways to test the behavior of the policy-informed model against a scenario: (1) regenerating the latest AI message, or (2) continuing the conversation by sending any message.}
\end{figure*}

Users can independently experiment with the behavior of the policy-informed model by \textit{regenerating responses} in the scenario sidebar (Fig. \ref{fig:testing-policy} \textbf{1}). Independent testing allows each user to focus on the specific concerns that drive their policy contributions, explore challenging boundary cases, and conduct stress-testing without group dynamics influencing their approach. Once they save the policy, they can also browse and compare responses generated by past policy versions. 

\begin{figure}[h]
    \centering
    \includegraphics[width=1\linewidth]{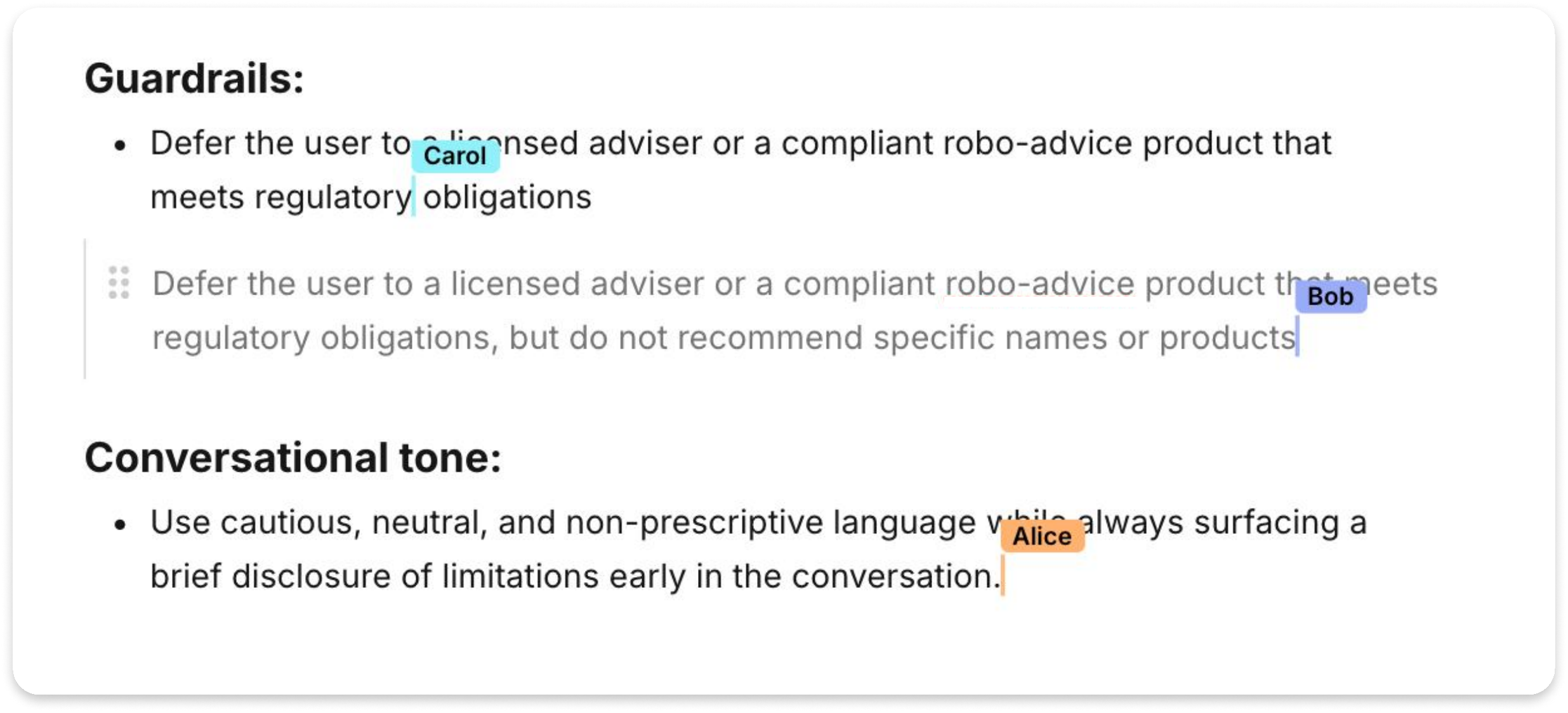}
    \caption{A drafting block (directly above ``Conversational tone'') can be added into the editor to draft experimental policies without affecting model behavior.}
    \label{fig:drafting-block}
    \Description{A drafting block that is in a slightly lighter shade of gray than the rest of the text can be added into the editor to draft experimental policies without affecting model behavior.}
\end{figure}

To propose edits in a non-disruptive way, users can add a \textbf{drafting block} (Fig. \ref{fig:drafting-block}) in the editor. Just like how a comment in a code editor is visible to a programmer but does not affect program behavior, content in the drafting block is visible to the group but is ignored by the model. After users review and reach consensus on changes, content in the drafting block can be integrated into the actual policy

To stress-test a policy, a user can \textit{extend} a scenario in the sidebar by continuing the existing conversation (Fig. \ref{fig:testing-policy} \textbf{2}). This offers an alternative way to experiment with policy-informed model behavior beyond regenerating a single message in a scenario.

Group members are not limited to only the scenarios initially provided to them. They may extend an existing scenario and add the extended version to the scenario gallery for others to view and extend further. They may also create a new scenario from scratch if none of the existing scenarios explore a particular behavior they want to test.

\begin{figure*}[h]
    \centering
    \includegraphics[width=1\linewidth]{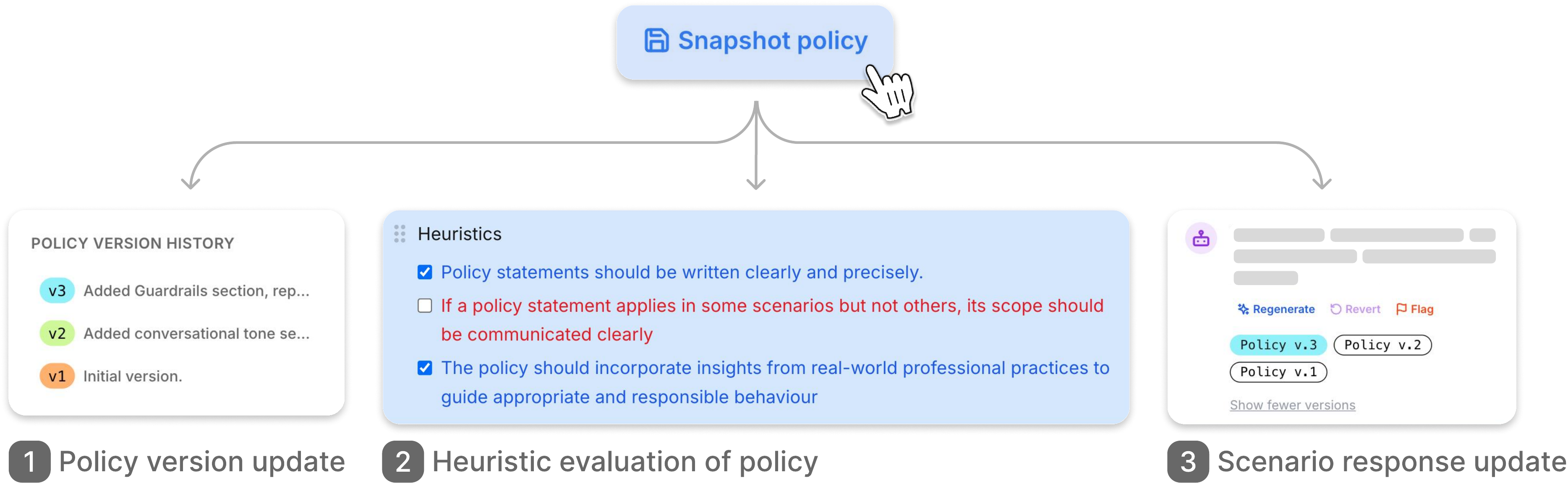}
    \caption{Upon saving the policy via the \textbf{Snapshot policy} button, \pp \textbf{(1)} adds the policy to the version history and generates a title summarizing key changes, \textbf{(2)} conducts an automated heuristic evaluation of the policy, and \textbf{(3)} updates the latest responses to all scenarios.}
    \Description{Upon saving the policy via the Snapshot policy button, PolicyPad (1) adds the policy to the version history and generates a title summarizing key changes, (2) conducts an automated heuristic evaluation of the policy and checks off heuristics that are satisfied, and (3) updates the latest responses to all scenarios.}
    \label{fig:snapshot-policy}
\end{figure*}

After a group has made meaningful edits to the policy, they can \textbf{save a new version}. After a user clicks the \textbf{[Snapshot policy]} button, \pp will add the current policy to the version history and regenerate the newest AI message for all scenarios using the policy-informed model (Fig. \ref{fig:snapshot-policy} \textbf{1\&3}).

\subsubsection{Scenario spotlights}
\begin{figure*}[h]
    \centering
    \includegraphics[width=1\linewidth]{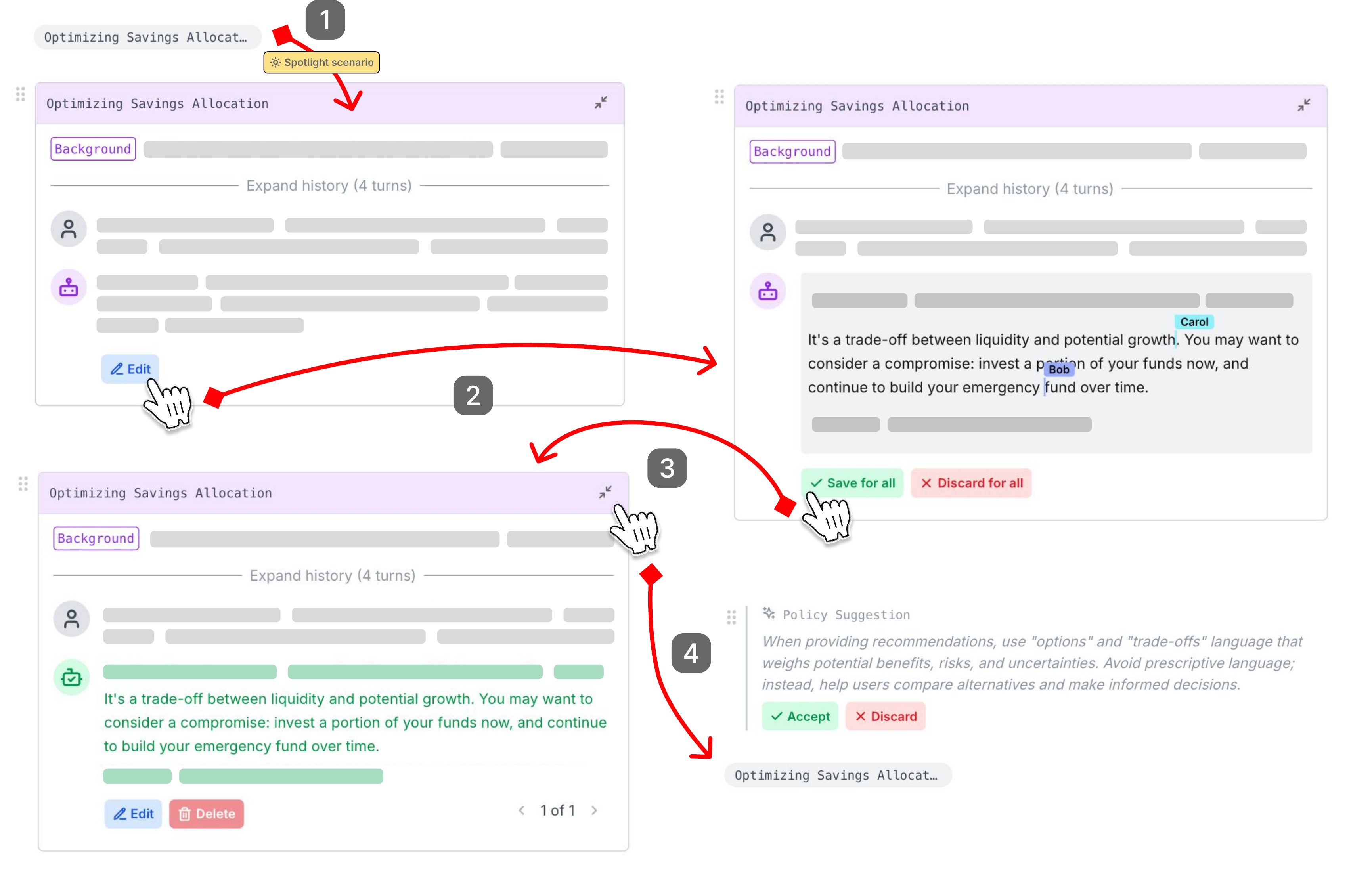}
    \caption{Workflow for spotlight scenarios. \textbf{(1)} A user can spotlight an interactive scenario widget to expand it into a card that everyone in the editor can view. \textbf{2)} The model's response can be edited collaboratively. After saving the edited response \textbf{(3)} and shrinking the spotlight scenario back into an interactive widget \textbf{(4)}, \pp automatically analyzes the edits in the context of the existing policy and heuristics to suggest a new policy statement.}
    \Description{UI workflow for spotlight scenarios. (1) A user can spotlight an interactive scenario widget to expand it into a card that everyone in the editor can view. (2) The model's response can be edited collaboratively. After saving the edited response (3) and shrinking the spotlight scenario back into an interactive widget (4), PolicyPad automatically analyzes the edits in the context of the existing policy and heuristics to suggest a new policy statement, which is inserted into the editor. The user may choose to accept or reject it.}
    \label{fig:spotlight-scenario}
\end{figure*}

Once a user has referenced a scenario in the editor, they can \textbf{spotlight} it to expand the interactive widget into a card UI that makes the full scenario visible to everyone (Fig. \ref{fig:spotlight-scenario}). Once spotlighted, the group can view, discuss, and even collaboratively edit the policy-informed model's response. Just like other text in the document, the scenario spotlight supports real-time collaboration for editing. After users are satisfied with their edits, any user can save the response. The old response remains easily accessible through a simple toggle. 

\pp then automatically analyzes the group's edits in the context of the current policy and heuristics. It uses a reasoning LLM (o4-mini) to \textbf{suggest a policy statement} designed to steer the model towards producing a response more similar to the edited version (Fig. \ref{fig:spotlight-scenario} \textbf{4}). If the user accepts the suggestion, it will be integrated into the policy. This alternative way of indirectly editing a policy through editing the model response is inspired by prior work on synthesizing principles from edits \cite{petridis2024constitutionmaker, louie2024roleplay}.

Once a group is finished with a scenario spotlight, a user can un-spotlight the scenario for everyone, shrinking the card back into a small, pill-shaped widget. 

\subsubsection{Heuristics editor and evaluator}

As a group expands and refines their policy, they may encounter additional considerations they would like to add as policy heuristics. For example, a policy may become increasingly riddled with domain-specific terms and acronyms unfamiliar to a layperson reading the policy, so the group may add a heuristic for clearly defining or explaining these terms. Since the likely audience of this policy is other people, factors like clarity and legibility are important to preserve.

Every time the policy is saved, \pp runs an automated heuristic evaluation using o4-mini to highlight any unsatisfied heuristics (Fig. \ref{fig:snapshot-policy} \textbf{2}). This automated evaluation is meant to draw attention to the heuristics and encourage discussion around them, rather than conclusively determining their fulfillment. Group members can easily override the automated decision if they agree on a different assessment. 

Users continue to engage in this iterative process of policy drafting and experimentation until the policy prototyping session concludes. 

\subsection{Technical Details}
\label{s:technical}
\pp is implemented as a web application built with React, TypeScript, and TipTap.\footnote{\url{https://tiptap.dev/}.} The real-time collaboration engine is supported via TipTap Cloud. \pp uses serverless functions to call the OpenAI and Together.ai APIs. The policy-informed model is an instance of Llama 3.3 70B Instruct Turbo\footnote{The ideal model for policy prototyping generates comparable responses to popular chatbot products (e.g., ChatGPT, Claude) but does not have an existing policy trained into it outside of basic guardrails. Llama 3.3 is an apt fit given its capable performance and bring-your-own safeguards setup (which we did not use). We also used Llama 3.3 instead of Llama 4 due to reports of the latter being narrowly optimized for specific benchmarks \cite{verge-llama4-gaming}.} hosted on Together.ai. The policy was fed into the model as a system prompt with some additional scaffolding to ensure the model followed it. We called GPT-4o for miscellaneous features that required light LLM processing (e.g., generating titles of new policy versions that capture key changes), and o4-mini for features that benefited from deeper reasoning (i.e. suggesting policy statements after response edits and automated heuristic evaluation). Our prompts are available in our Supplementary Materials.

\section{Evaluation Study}
To evaluate \pp, we ran a series of group-based, within-subjects studies with 22 domain experts from two domains (10 from mental health, 12 from law). Our goal was to determine how the design decisions made for \pp enhanced the policy prototyping experience. We also evaluate the outputs of \pp by analyzing the policies created by experts to determine their novelty with respect to established LLM  policies like Claude's Constitution \cite{claude-constitution} and OpenAI's Model Spec \cite{oai-model-spec}.
Thus, we asked the following research questions:

\begin{itemize}
    \item[\textbf{RQ1:}] How did the individual components of policy prototyping supported by \pp (rapid policy iteration, heuristic evaluation, interaction with scenarios, real-time collaboration) aid expert-driven policy design in practice?
    \item[\textbf{RQ2:}] To what extent are the insights in experts' policies created through \pp  novel compared to existing, publicly available LLM policies? 
\end{itemize}

We selected mental health and law as our domains because they are regulated, high-stakes domains for which AI use has been increasing but contested\footnote{OpenAI CEO Sam Altman observed that the younger generation uses it as a \textit{``therapist, a life coach [...] asking, `What should I do?'''} despite the absence of confidentiality protections that govern human attorneys and therapists~\cite{techcrunch}.} \cite{lamparth2025moving, moore2025expressing, cheong2024not, nyt-ai-lawyers}. Crafting responsible LLM policies is therefore critical for ensuring users' safety and well-being. The two domains are also distinct enough for us to observe how approaches to policy prototyping and the resulting policies can differ across domains. For each domain, we randomly sampled scenarios from datasets compiled by prior work in mental health \cite{lamparth2025moving} and law \cite{cheong2024not}. More details about scenario construction can be found in Section \ref{s:eval-starting-materials}. 

\subsection{Participants and Setup}
We recruited 22 domain experts (Table \ref{t:eval-participants}) through our personal connections, university mailing lists, professional Slack channels, and snowball sampling. Among mental health experts ($n=10$, 7 female, 3 male), the average years of practical experience\footnote{We define ``practical experience'' as conducting client-facing work at a clinical or legal organization.} held by each expert was \textbf{10.5} (min 3, max 25). Among legal experts ($n=12$, 5 female, 7 male), the same figure was \textbf{5.6} (min 2, max 13). Five of the mental health experts also participated in our earlier observational study. 

We organized experts into small groups of 2--4 (median = 3). We observed during our formative study that groups of around three experts struck an ideal balance between creating a collaborative atmosphere and allowing room for meaningful individual contributions. Full participant and group details are available in Appendix \ref{a:eval-participants}. Half the groups (4 of 8) contained participants who already knew each other from professional contexts. We did not observe this to impact the quality of discussions nor policies prototyped.

We facilitated policy prototyping sessions by providing (with order counterbalanced) \pp and a baseline system to each group, followed by a brief exit interview. The total length of the study was 90 minutes. Participants were compensated \$150 USD in their choice of cash or a gift card upon completing the study. All studies were recorded and transcribed. This study was classified as exempt by the University of Washington IRB. 

\subsubsection{Tasks}
We assigned each group two policy prototyping tasks, corresponding to distinct sections of a policy. One addressed the \textbf{conversational tone}---guidelines for how the model communicates with users. The other addressed \textbf{guardrails}---hard-and-fast rules that constrain model behavior for safety and legal compliance. Tasks were counterbalanced by system condition and task order in a 2×2 factorial design.

\subsubsection{Baseline system}
\label{s:baseline}
Our baseline system implemented a simplified policy prototyping workflow that consisted only of iterative policy drafting and experimentation with a policy-informed model identical to the one used in \pp. It used the same collaborative policy editor as \pp, but did not include built-in support for interactive scenarios (i.e, the scenario sidebar, scenario widgets, spotlight scenarios) nor heuristic evaluation. It resembled a more polished version of the first prototype used in \autoref{s:system:co-design-session}: an editor with a policy-informed model in the sidebar (Fig. \ref{fig:pp-v1}). We chose this baseline because it mimicked the system design of many existing ``copilot'' systems that embeds a chatbot in a sidebar alongside a document environment (e.g., Copilot for Microsoft or Gemini in Google Docs) as well as policy authoring setups used in prior literature \cite{cheong2024not}.

\subsubsection{Starting materials}
\label{s:eval-starting-materials}
We prepared scenarios, heuristics, and a small amount of starter text for the policy. For each domain, the first author crafted 10 scenarios that represent a realistic conversation between a human user and an AI chatbot, using the same Llama 3.3 instance as \pp to generate the responses. To maximize realism without access to product interaction logs, we sourced topics and language from datasets containing realistic questions in our domains of interest: MENTAT from Lamparth et al. \cite{lamparth2025moving} for mental health, and \texttt{r/legaladvice}-style cases from Cheong et al. \cite{cheong2024not} for law. We varied the length of scenarios to be 1--5 conversational turns. For multi-turn scenarios, follow-up user messages were written by the first author. For each group of experts, we randomly sampled half the scenarios (5) to assign to the first task, and the rest were assigned to the second.\footnote{We note that seeding a policy prototyping session with a different set of scenarios may change its policy outcomes; we discuss how this can be leveraged favourably in this in Section \ref{s:limitations-fw}.} In the system condition, scenarios were loaded directly into the system. In the baseline condition, the first user messages across the 5 scenarios were copied into a Google Doc and shared with participants upon request.\footnote{In the baseline condition, we gave participants the option of starting with or without first browsing these user messages.}

Besides scenarios, we provided three basic heuristics to encourage clear and precise policy writing that draws from real-world professional practices. We also provided an Objectives section in the policy as examples of policy statements, drawn from objectives in OpenAI's Model Spec \cite{oai-model-spec}. The full starter heuristics and policy are available in Appendix \ref{a:starter-material}.

\subsection{Procedure}
The studies proceeded as follows:
\begin{itemize}
    \item \textit{Introduction [5 mins]}: The facilitator introduced the study and agenda, and participants each introduced themselves to each other.
    \item \textit{Task 1 [30 mins if baseline, 40 mins if system]}: The facilitator oversaw a minimally structured policy prototyping session for either the conversational tone or guardrails. In the baseline condition, 5 minutes were used for a brief demo. This included some time for participants to try out features for themselves. In the system condition, this demo period lasted 15 minutes due to the additional features. The time dedicated to policy prototyping was 25 minutes in both conditions.
    \item \textit{Task 2 [30 mins if baseline, 40 mins if system]}: The procedure for Task 1 was repeated for a different task in a different system condition. 
    \item \textit{Exit interview [15 mins]}: The facilitator asked each participant to reflect on their experiences across the system and baseline systems. Participants were also asked to share what they were most excited and concerned about regarding AI use in their domains. 
    \item \textit{Post-study survey}: Group members swapped policies with another group in their domain and rated the policies on 5-point Likert scale questions (see Appendix \ref{a:policy-likert-qs}). 
\end{itemize}

Throughout the study, the facilitator (first author) ensured discussions between participants went smoothly and followed up on specific points when the conversation died down, but otherwise tried to let participants drive the session. 

\subsection{Data Analysis}

\subsubsection{Thematic analysis (RQ1)}
The first author qualitatively coded the study transcripts using reflexive thematic analysis \cite{braun2006using, braun2019reflecting}. An initial deductive pass isolated specific parts of the transcript that were highly relevant to each research question, followed by one or more inductive passes to surface themes organically. This analysis was augmented by short memos the facilitator wrote upon concluding each study, summarizing key events and noteworthy insights from each session.

\subsubsection{Policy novelty analysis (RQ2)}
\label{s:method-novelty-analysis}
To analyze the novelty of policies---whether they contribute new perspectives, ideas, considerations, dependencies, or approaches to existing policies---we gathered experts' policy statements from all sessions and evaluated each against publicly available policies for guiding responsible model behavior (the ``existing set''). We combined all policies from OpenAI's Model Spec \cite{oai-model-spec}, Claude's Constitution \cite{claude-constitution}, and principles derived from workshops with legal experts in prior work \cite{cheong2024not}, to represent the set of existing policies.\footnote{Excluding policies specifically targeted at moderating hate speech and disturbing content (e.g., \cite{lam2024ai}), as they are orthogonal to the policies we focus on in the prototyping sessions.} 

Rather than rely on direct human coding of novelty between policies, we opted for a joint human-AI approach where we made use of LLMs to first identify portions that were \textit{likely to be novel} before having human annotators review and make the final novelty determinations. 
Specifically, we followed this procedure: 
\begin{enumerate}
    \item For each expert-written policy statement, we used 3 prompts with varying definitions of novelty\footnote{Executed on GPT-4.1. Prompts provided in Supplementary Materials.} to make binary novelty decisions against the existing policies. Our prompts also required the LLM to generate a justification for its decision. 
    \item Any policies that were not unanimously determined to be novel in all 3 prompt evaluations were considered not novel. For the remaining policies, we further prompted the model to retrieve relevant quotes from the existing policies to be used as context for human evaluation.\footnote{Also through GPT-4.1.} Retrieved quotes were examined by the first author alongside the full existing set to catch for hallucinations and any missed quotes. 
    \item Finally, two human annotators (members of our research team) reviewed the list of policies and quotes to make a final novelty determination. Annotations were first done independently, with an initial Cohen's Kappa of 0.41. Disagreements primarily arose from differences in the interpretation of novelty; they were resolved via a round of discussion. If annotators failed to reach a consensus (which was the case for 2 policy statements), the policy statement was considered not novel by default. 
\end{enumerate}

While the reliability of LLMs for making content judgments has been called into question by recent work \cite{thakur-etal-2025-judging, chehbouni2025neither,szymanski2025limitations,kim2025correlated, krumdick2025no,chehbouni2025neither}, we structured our evaluation process to minimize the potential impacts of these factors through deliberate prompt design. 
Specifically, we attempted to control for model sensitivity to prompts by using multiple variations of prompts for initial evaluation. Additionally, we incorporated existing policies from prior work \cite{oai-model-spec, claude-constitution, cheong2024not} into the prompt and request justifications and quote extraction to identify possible model hallucinations. Finally, we ensured the final novelty determination is done by human annotators. 
Overall, we believe that this process should yield a \textit{conservative} determination of novelty, while also ameliorating challenges around human attention during review and comparison of exceptionally long texts. Our prompts used in this evaluation are available in our Supplementary Materials.

\section{Findings}

\subsection{Design Decisions in \pp Fostered Collaboration During Policy Prototyping (RQ1)}
\label{findings:rq1-system-pros}

\subsubsection{Heuristics built common ground and inspired richer policies \textbf{(system only)}}
\label{findings:rq1-heuristics}
Participants generally agreed that having heuristics as part of the policy prototyping process helped develop common ground for the group. P3 found that heuristics helped the group align on \textit{``the spirit of what we were doing''} and ensure \textit{``we're on the same page about the purpose of the policy.''} P19 had a similar experience: \textit{``[the heuristics] set the underlying tone for how the policy is supposed to function.''} P5 thought the heuristics gave \textit{``an idea of what sorts of [policy statements] would work best,''} while P4 agreed, finding that heuristics offered \textit{``more specific guidance''} for drafting policies around edge-cases in model behavior. P7 appreciated heuristics as \textit{``a constant reminder of the guidelines,''} but recognized that because domain experts are already well-acquainted with many of these guidelines, heuristics might be even more useful for developers who are refining the policy and integrating them into models.

Besides serving as guidelines, heuristics also served as entry points for deeper discussion on key policy topics. The starter heuristic on incorporating real-world professional practices into the policy initiated discussions in MH01, MH02, and MH04 about \textit{motivational interviewing} (MI), a foundational technique for therapists. Experts then incorporated various aspects of MI into the policy, such as encouraging \textit{``summaries of conversation when appropriate''} (MH01), and \textit{``Repeating or paraphrasing what [the user] is saying''} (MH02). MH02 and MH03 brought up \textit{limits of confidentiality}---when a mental health expert is legally or ethically required to break confidentiality to share client information, even though expert-client conversations are otherwise private. MH02 brainstormed situations in which experts needed to break confidentiality (e.g., \textit{``immediate risk of harm to oneself or others; suicidal thoughts, urges, or behaviors; presence or risk of non-suicidal self injury''}) and added a policy to \textit{``avoid using MI''} in those situations. MH03 agreed that models, just like when experts work with clients, should provide a disclaimer early on in the conversation of conditions under which confidentiality will be broken and remind the user that \textit{``[the conversation] is not a confidential setting''} when those conditions are triggered.  

Interestingly, a couple groups of legal experts used the heuristic to debate whether real-world practices for lawyers and other legal professionals should even apply to AI, since they clearly established that AI does not have the same legal status as human lawyers. P17 in L02 shared that while \textit{``we lawyers do have rules of professional responsibility that we need to adhere to, they don't apply to non-lawyers''} and suggested removing that heuristic. P20 in L03 echoed that sentiment: \textit{``Those ethical rules of lawyers do not apply to AI systems.''} Their group member, P19, agreed, and noted that \textit{``if ethical rules of lawyers did apply, then the AI model cannot even begin to suggest answers.''} In general, this is an important point of distinction between the mental health and legal policies; we unpack this in more detail in our paper's Discussion.

We observed that groups did not initially modify the starter heuristics provided to them, but some added more heuristics as they worked on their policy. For example, group MH02 realized their policy contained some mental health-specific concepts that needed to be explained to the facilitator, and added a couple heuristics to \textit{``Give illustrative examples for concepts and terms''} and \textit{``Give definitions for jargon and technical terms.''} Similarly, L03 added a heuristic to \textit{``Clearly explain or define legal terminology.''} As their policy got longer, L02 added a heuristic to ensure \textit{``No policy statements should conflict with each other.''}

\subsubsection{Spotlight scenarios improved collaborative dynamics and provided valuable writing support \textbf{(system only)}}
Participants found the ability to bring scenarios into the editor, spotlight it, and collaboratively edit its response to be valuable features in \pp. In general, we observed a general pattern where participants referencing scenarios in their discussions and then referenced them in the editor for others to view. P22 said the ability to \textit{``input the scenarios into the editor helped with brainstorming and being able to point to specific parts of a response we either found helpful or that we thought needed to be changed.''} P5 agreed and thought that the utility of spotlight scenarios could scale with the number of collaborators: \textit{``[the spotlight] would be really nice for larger groups of people contributing, being able to look at [the scenario] together.''}  P1 thought spotlight scenarios can be useful for facilitating asynchronous collaboration as well: \textit{``I see [P3] has already edited this side of things. I can hop right back in and draft out a version of the [policy] and stress test it.''} After using the interactive scenarios, P11 thought that \textit{``for a collaborative effort, [\pp]'s really nice. [The baseline] felt more like a personal tool.''}

Indeed, in the baseline condition, whether it came before or after the system condition, experts were finding makeshift ways to accomplish what scenario spotlighting are designed for. For instance, P4 asked to share their screen so everyone could view the policy-informed model response they generated. Similarly, when P8 was pointing to a specific aspect of a generated response, their groupmate P7 asked: \textit{``Is that something you can share so we can all see it, so we just work off of that one?''} In both cases, experts improvised a solution by using a \textbf{drafting block} to share content from scenarios in the editor without impacting the policy.

Participants also expressed appreciation of the ability to edit the response and receive a system-generated policy suggestion. 100\% of the policy suggestions were accepted in our studies. P17 shared that the suggestions were helpful in articulating their thoughts: \textit{``sometimes it's difficult to put your thoughts into words, and the [suggestions] are helping you with that.''} P10 agreed, saying that \textit{``the ability to pull that response in, edit it, and have a generated guardrail could be a huge time saver.''} They saw as a way of removing the need for low-level wordsmithing: \textit{``We don't need to edit that response perfectly, but if we can make it clear what our priorities are, and then see if the AI gets our nuance, that's pretty incredible.''} We also observed that in MH04, P9 and P10's policy drafting began slowly, but accelerated considerably when they received policy suggestions that inspired more ideas. P7 shared why they accepted policy suggestions even when they seemed imperfect: \textit{``I liked the policy [suggestions], even if they weren't necessarily dead on. It gave us ideas for other [policy statements].''}

\subsubsection{Experimentation with a policy-informed model directly informed policy edits \textbf{(system \& baseline)}}
In both the system and baseline conditions, we observed how quick and iterative experimentation with the policy-informed model\footnote{Recall from Section \ref{s:baseline} that the baseline uses the same model as \pp.} benefited the policy prototyping process. Experts easily surfaced specific model behavior that could be addressed with the policy, such as when the model was overstepping its role, such as making a judgment about \textit{``whether a risk is worth or not worth taking''} (P13). Experts could then draft the policies and immediately observe the impact their edits had on the response, either by clicking the [Regenerate] for quick testing or taking a snapshot of the policy and updating all responses to all scenarios at once. As experts critically evaluated the responses for common behavior they targeted in the policy---such as judgmental language (L01, L02, L04, MH01, MH02, MH03), eliciting necessary information from users in order to provide a responsible answer (L01, L02, L03, MH01, MH02, MH04), and the inclusion of disclaimers (all groups)---they could qualitatively observe clear improvements. For example, at the end of their session, P19 confirmed that \textit{``I see everything we've discussed being implemented, and [the model] still manages to give a fair amount of information, so that's good.''} 

\subsubsection{Real-time collaboration amplified other benefits \textbf{(system \& baseline)}}
Participants actively engaged with each other during the sessions---seeking and providing peer feedback, discussing nuances and complexities of model behavior, sharing insights from their own professional experiences, and more. This engagement benefited all components of the policy prototyping workflow. 

They provided input for and helped edit responses when a group member put a scenario on spotlight. They shared explorations of edge cases in model behavior with the group to patch gaps in the policy and identified high-risk scenarios to focus their discussions. Overall, real-time collaboration amplified participants' abilities to draw upon their expertise, challenge assumptions, and iteratively refine policies.

We observe that participants tended to agree with others in their group and rarely challenged or pushed back directly on others' input. This may be due to a desire to appear diplomatic and accommodating, especially when working with new collaborators.

\subsection{Experts Prototyped More Novel Policies in \pp Than the Baseline (RQ2)}
\label{findings:rq2-novelty}

\begin{figure*}[h]
    \centering
    \includegraphics[width=1\linewidth]{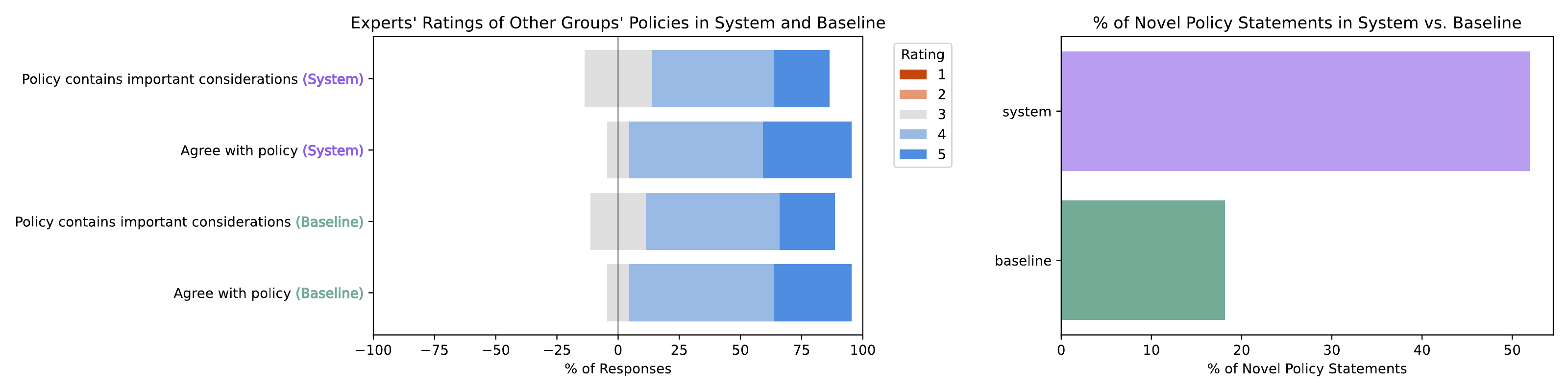}
    \caption{Comparison of Likert scale responses to policies prototyped in the system vs. baseline \textbf{(left)} and the novelty of policy statements prototyped in the system vs. baseline \textbf{(right)}.}
    \Description{Two bar charts side by side. On the left, a stacked bar chart with 5-point Likert scale ratings of policies created in the system and baseline conditions. The conditions are comaprable and have a median rating of 4. On the right, a chart showing that the percentage of novel policies created in the system is a bit above 50 percent, while it is just below 20 percent in the baseline.}
    \label{fig:quant-results}
\end{figure*}

\subsubsection{Quantitative results}
Our novelty analysis (Section \ref{s:method-novelty-analysis}) revealed that experts prototyped more novel polices using \pp compared to the baseline (Fig. \ref{fig:quant-results} right). \textbf{51.9\%} of the policy statements drafted in \pp were considered novel, compared to \textbf{18.2\%} from our baseline. Looking at raw numbers, the number of novel policy statements from \pp was \textbf{4 times} that of the baseline (40 vs. 10). 

Outside of novelty, we found the policies from the two systems to be comparable (Fig. \ref{fig:quant-results} left). After the study, experts rated policies from another group within their domain along two dimensions: the extent to which 1) the policy contained \textit{important considerations} of AI behavior within their domain, and 2) they \textit{agree} with the policy. Wilcoxon signed-rank tests on Likert data showed no significant difference ($W=6.0; p=0.65; M_{system}=M_{baseline}=4$ for important considerations, $W=20.0; p=0.74; M_{system}=M_{baseline}=4$ for agreement). This suggests that experts generally viewed policies from other groups favorably, and that novelty was the main differentiator between policies across the two systems. We suspect the high positive bias may stem from collective alignment on priorities when serving clients, which in turn may be due to some standardization in professional training (e.g., motivational interviewing in mental health).

\subsubsection{Qualitative results}
\label{s:qual-results}
We conducted a qualitative analysis of the novel policies to understand what exactly was novel about them. We identified three main sources of novelty.

First, experts' policies offered more insight into \textbf{specific circumstances under which the model should defer the user to a human expert}. MH01 noted that while the model can provide empathy and reassurance, as soon as indications of \textit{``behavioral interventions (such as behavioral activation [for depression], exposure and response prevention [for OCD], or prolonged exposure [for PTSD])''} arise, the model should defer to a human therapist. L03 shared that the users' desires to share confidential information with the model is a good indicator of whether the model should defer to a legal expert: \textit{``If the user indicates that they want or need to provide confidential information, there may be privileged information involved. If the conversation contains privileged information, always defer the conversation to a legal expert.''} L04 summarized their perspectives on this issue as: \textit{``[The model should] answer 'what can I do' questions and defer 'what should I do' questions to a lawyer.''} Generally, current models' lack of awareness of when to defer to human experts is a significant safety concern \cite{nyt-chatgpt-suicide}, and experts' policies have potential to improve this awareness. 

Second, experts' policies provided \textbf{specific procedural guidelines} that existing policies lack or do not specify in much detail. MH01 and MH02 both provided guidelines for motivational interviewing in their policies, but noted that the technique should not be used in high-risk situations. In such situations, MH04 emphasized the model should be much more succinct and direct when supplying crisis response information: \textit{``When a user indicates a high level of distress, or [when] crisis services or hotlines are required, provide them succinctly and without much additional text.''} MH03, L02, and L03 all instructed the model to disclose limits to or the lack of confidentiality early in the conversation. MH03 specifically stated that after this disclosure, the model should then \textit{``Ask users if they have questions about confidentiality limits,''} reflecting a procedure in their own mental health practice.

Finally, experts' policies recommended the model to be \textbf{more proactive about seeking key information at the start of the conversation}. Experts also considered it irresponsible for the model to assist users when lacking key information, such as legal jurisdictions. L03 wrote that the model should \textit{``require the user to indicate their jurisdiction before providing full responses.''} Similarly, L02 recommended the model to avoid providing assistance if it cannot elicit \textit{``essential case details (such as date of offense, location, and current legal status) necessary to tailor legal guidance.''} MH01 and MH04 both wanted the model to conduct a more thorough risk assessment prior to engaging deeply with the user. MH01 recommended that the risk assessment include \textit{``asking how problems or challenges have been addressed (or not addressed) before, how long the problem has persisted, and how distressing/problematic the user finds the current situation.''} These policies can help augment existing technical efforts in improving LLMs' abilities to elicit user information to improve their quality of assistance \cite{andukuri2024star, li2024mediq}.

\section{Discussion}

\subsection{Practical Deployment Considerations}
Results from our policy prototyping sessions showed that experts can contribute novel policies to shape model behavior in domain-specific, high-stakes interactions with users. As these interactions become more commonplace, it is crucial for users' safety that model behavior is vetted, scrutinized, and refined by expert input. Policy prototyping is one promising avenue for this, but numerous considerations for widespread adoption remain. These include resource constraints, time commitment, learning effort, and impacts on experts' own careers. We now discuss these considerations and some possible ways forward.

\subsubsection{Time and resource intensity}
First, policy prototyping can be time- and resource-intensive. Our sessions were 90 minutes in length and could have easily been longer if not for scheduling constraints. Our sessions were one-off events, but LLM policy design can and should benefit from \textit{sustained} expert engagement \cite{oai-expert-input, Walsh2025VirologistOA}. Challenges associated with time and resource intensity may be alleviated if policy prototyping was more seamlessly integrated into experts' professional practices and did not cold-start like it did in our work. After all, academic or industry labs are not the only ones motivated to design better LLM policies---expert communities \textit{themselves} are too. As AI diffuses into their domains, they are incentivized to take bottom-up approaches\footnote{Assuming that model developers are open-minded about integrating at least some of experts' suggestions.} to shape model behavior, due to its ability to impact the behavior and actions of clients \cite{nyt-chatgpt-psychosis, cheong2024not}. Indeed, the American Psychological Association (APA) has started to track the use and integration of AI into mental health practice as a key industry trend \cite{Abrams2025apa}. APA and similar organizations may thus be interested in hosting policy prototyping sessions at gatherings or conferences, potentially in partnership with AI developers, as a way of actively engaging the community to shape AI diffusion in their domain. Additionally, local chapters may have the capacity to host sessions more regularly as part of ongoing discussions. 

\subsubsection{Expert displacement}
Relatedly, it is reasonable to think that experts' efforts in policy prototyping could result in AI taking over large portions or even all of their work. A therapist, for example, might design a policy that faithfully reproduces how they interact with clients, only to find that the AI could then offer a cheaper substitute of their services. This possibility highlights an important benefit of policy prototyping: \textit{it equips experts with a voice to shape how AI can impact their work.} Instead of reproducing therapeutic behavior, mental health experts can encode guardrails that push the model away from emulating them directly. Indeed, several groups adopted this approach in our study (see Appendix \ref{a:disagreement}).
As long as there are incentives for model developers to adopt policies created from third-party input, which prior work has shown to be the case \cite{huang2024collective, openai-collective-alignment}, policy prototyping offers experts more agency to guide how AI diffuses throughout their domains. 

\subsubsection{Organizations for facilitation}
Finally, who can facilitate policy prototyping sessions? In this work, the sessions are facilitated by an academic lab interested in studying expert-informed LLM policy design. Being affiliated with a university allowed us to take advantage of being in close proximity to---or already connected with---experts (e.g., in academic departments, the medical school, the law school, etc.). However, in practice, we see a wide range of organizations with diverse resource profiles that can collaborate with each other for even more effective facilitation. For example, non-profits interested in AI safety, such as METR\footnote{\url{https://metr.org/}} and the AI \& Democracy Foundation,\footnote{\url{https://aidemocracyfoundation.org/}} alongside organizations advocating for responsible AI more generally (e.g., Partnership on AI,\footnote{\url{https://partnershiponai.org/}} Ada Lovelace Institute\footnote{\url{https://www.adalovelaceinstitute.org/}}), can set agendas for policy prototyping that target emerging harms and underrepresented stakeholder concerns. National AI safety centers (e.g, UK AISI,\footnote{\url{https://www.aisi.gov.uk/}} US CAISI\footnote{\url{https://www.nist.gov/caisi}}) can guide policy design efforts to prioritize legitimacy and trust in the public interest. Safety and alignment teams within frontier model development companies (e.g., Anthropic, OpenAI) can leverage resources for expert recruitment and contribute technical tooling.  Ultimately, effective facilitation will likely require combining these strengths across institutions rather than relying on any single actor.

\subsection{Scaling Up Policy Prototyping for Participatory Model Behavior Design}
Our work focused on engaging small groups of domain experts in deliberating on and actively creating policies for model behavior. We see potential in broadening our approach to beyond collaborative design with experts. This is valuable when democratic participation is prioritized over narrowly defined expertise, or for topics where the notion of ``expertise'' is blurry (e.g., personal information management). To expand beyond experts and reach towards more \textit{participatory} visions of AI development with diverse stakeholder groups \cite{delgado2023participatory, suresh2024participation, meta-community-forums, barnett2025envisioning, kallina2025mapping}, efforts to scale up policy prototyping---including tool ecosystems that do not only consist of \pp---will be required. 

A challenge that will inevitably arise with increasing scale is resolving disagreement. We have already started to see this challenge emerging within our small-scale groups---expert groups within and across domains did not always agree on how to design the policy. Within \textit{mental health groups}, there was some disagreement over 1) whether the model should act like a therapist, and 2) the appropriate conversational tone before a proper assessment of the user is made. Within \textit{legal groups}, experts disagreed over whether the model should suggest action items for the user---a behavior that was found to be unanimously criticized by legal experts in prior work \cite{cheong2024not}. \textit{Across the domains}, disagreements arose over whether the model should, under any circumstances, attempt to mimic a human professional. More details about these disagreements can be found in Appendix \ref{a:disagreement}.

Prior work has attempted to resolve disagreements on AI behavior at scale through multiple rounds of voting and asynchronous deliberation \cite{huang2024collective}. However, asynchronous collaboration methods rarely uncover the same richness and nuances of disagreements that we observed in our synchronous, deliberative policy prototyping sessions. To scale policy prototyping to include more stakeholders while preserving the quality of discussions, we may draw inspiration from multi-stage or tiered citizens' assemblies (also known as ``mini-publics'') \cite{we-the-citizens, global-assembly, meta-community-forums}. These systems aggregate deliberations from parallel, local assemblies into regional or (inter)national assemblies for producing recommendations. For policy prototyping, parallel deliberations may be held on a regular basis with groups of individuals whose personal and professional lives are impacted by model behavior so they have more agency to shape it (e.g., parents who are concerned about their children interacting with AI companions, workers who are expected to use AI tools to boost their productivity). The resulting policies can then be aggregated with policies prototyped by domain experts in parallel. Aggregations can also happen across geographic regions to incorporate pluralistic cultural and social perspectives. 

New suites of tooling and infrastructure will likely be needed to facilitate policy prototyping at scale. At the deliberation level, tools like \pp can help. At the aggregation level, civic technologies like Pol.is \cite{small2021polis} are more suitable. We encourage future work to explore different combinations of collaborative affordances and interaction paradigms to build new tool suites for scalable policy prototyping.

\subsection{Situating Policy Prototyping Within AI Alignment}

We proposed policy prototyping as a practice through which small groups of policy designers can collaboratively design LLM policies. Where can this practice fit within the broader AI alignment pipeline?

A prerequisite for policy prototyping is an instruction-tuned model with behaviors representative of what users will experience in the wild. This is most commonly a frontier model---a model with frontier performance on popular benchmarks---as they are commonly integrated into user-facing applications. Thus, we envision policy prototyping taking place \textit{after} fundamental alignment and safety efforts (e.g., instruction tuning, RLHF, implementing basic safety guardrails and classifiers, automated red-teaming). However, policy prototyping should come \textit{before} more later-stage or sophisticated alignment efforts that benefit from or even require a policy (e.g., deliberative alignment \cite{guan2024deliberative}, manual red-teaming \cite{ahmad2025openai}). If a developer has an existing policy, policy prototyping can reveal nuances, inconsistencies, and areas needing refinement before the developer commits significant resources to align the model with it. If the developer does not yet have a policy, policy prototyping can help start one.

We also note that LLM policies are continuously evolving artifacts, rather than static ones \cite{oai-model-spec}. Model developers may thus find it helpful to \textit{co-evolve} the policy and alignment strategies. This can involve policy prototyping sessions with experts on a regular basis to seek input on top-of-mind concerns based on insights from usage telemetry.

\section{Limitations and Future Work}
\label{s:limitations-fw}
All our participants except for two legal experts were based in the U.S.. The perspectives integrated into our policies are thus heavily influenced by the American mental health and legal systems, and may not generalize to other countries. Before integrating these policies into AI systems that serve a global userbase, future work should augment and contrast our policies with perspectives of non-US and non-Western experts.

The selection of scenarios for policy prototyping can alter policy outcomes by steering discussions towards topics depicted in the scenarios. We selected scenarios based on realistic tasks experts in mental health and law might tackle in their day-to-day work, but our scenarios may not achieve adequate coverage over the diversity of scenarios experts actually encounter. We also manually composed responses to the LLM in multi-turn scenarios, which may not reflect real-world responses from users asking about mental health or legal questions. Further, we only started with 5 scenarios per session to fit the study within the 90-minute time limit.  Future work can explore the optimal number of scenarios for a certain group size, as well as dedicating time for group members to author scenarios prior to prototyping the policy to further improve realism. 

Researchers running future policy prototyping sessions may be interested in potential modifications of our setup. We recommend exploring three modifications. First, the facilitator for a workshop holds a major role and can influence the results with their facilitation style. The first author facilitated all sessions in our study for consistency, but future work can experiment with different facilitation styles to determine which are more effective. Second, the starting scenarios can focus on a particular themes or issue within a domain for more targeted policy design. For example, due to recent high-profile cases of AI-driven psychosis \cite{nyt-chatgpt-psychosis, time-chatbot-psychosis, nyt-psychosis-transcript}, scenarios can draw from real transcripts of psychosis-inducing conversations \cite{nyt-chatgpt-psychosis} rather than our random sampling approach. Third, the sessions can be scaffolded with taxonomies of concepts within a specific domain. Prior work relied on concepts as a central ingredient in policy design \cite{lam2024ai}. While concepts were not fundamental to our work, including them may benefit future sessions. 

Finally, experts agreed the model should elicit key information from users before providing assistance (Section \ref{s:qual-results}). Effective elicitation requires the model to \textit{reason about missing information}. While it is promising that LLMs' reasoning capabilities have improved significantly in recent months, improvements have primarily focused on verifiable domains like math and coding \cite{muennighoff2025s1}, and it is unclear whether these improvements translate to more effective information elicitation. Future work can empirically investigate this and develop techniques for models to reason about missing information in contextual, human-centered ways.

\section{Conclusion}
In this work, we asked: \textit{How can domain experts be meaningfully involved in designing LLM policies as a means of actively shaping responsible model behavior?} In response, we introduced \pp, an interactive system for small groups to engage in LLM policy prototyping---a practice that draws upon UX prototyping methods to enable collaborative drafting, testing, and rapid iteration of LLM policies in real time. We conceptualized LLM policy prototyping and motivated the design of \pp through a 15-week observational study with 9 mental health experts. We then evaluated \pp through 8 policy prototyping sessions with 22 experts in mental health and law. We found that \pp fostered a collaborative and productive dynamic for policy prototyping and led to the creation of more novel policies compared to a baseline. Areas of novelty covered important considerations for model behavior, such as when to defer to human experts, specific procedures for emergency situations, and eliciting missing information needed to responsibly provide assistance. We hope future work will extend our contributions and continue co-designing policies with experts and laypeople alike to improve the safety and responsibility of advanced AI.

\begin{acks}
We extend a warm thanks to all our participants for their time, expertise, and thoughtful engagement with our studies. We also thank our anonymous reviewers for feedback on our manuscript. Finally, we thank Tyna Eloundou, Teddy Lee, and others in OpenAI's Democratic Inputs to AI grant program for their support and feedback on this project.
\end{acks}

\bibliographystyle{ACM-Reference-Format}
\bibliography{refs}

@article{ahmad2025openai,
  title        = {OpenAI's Approach to External Red Teaming for AI Models and Systems},
  author       = {Ahmad, Lama and Agarwal, Sandhini and Lampe, Michael and Mishkin, Pamela},
  year         = {2025},
  journal      = {arXiv preprint arXiv:2503.16431}
}

@article{anderljung2023frontier,
  title        = {Frontier AI regulation: Managing emerging risks to public safety},
  author       = {Anderljung, Markus and Barnhart, Joslyn and Korinek, Anton and Leung, Jade and O'Keefe, Cullen and Whittlestone, Jess and Avin, Shahar and Brundage, Miles and Bullock, Justin and Cass-Beggs, Duncan and others},
  year         = {2023},
  journal      = {arXiv preprint arXiv:2307.03718}
}

@article{andersen1999scenario,
  title        = {Scenario workshops and consensus conferences: towards more democratic decision-making},
  author       = {Andersen, Ida-Elisabeth and J{\ae}ger, Birgit},
  year         = {1999},
  journal      = {Science and public policy},
  publisher    = {Beech Tree Publishing},
  volume       = {26},
  number       = {5},
  pages        = {331--340}
}

@article{andukuri2024star,
  title        = {Star-gate: Teaching language models to ask clarifying questions},
  author       = {Andukuri, Chinmaya and Fr{\"a}nken, Jan-Philipp and Gerstenberg, Tobias and Goodman, Noah D},
  year         = {2024},
  journal      = {arXiv preprint arXiv:2403.19154}
}

@misc{quicksey-policy-prototypes,
  title        = {{Policy Prototypes: How designers and policy practitioners can use prototypes to get feedback and iterate on policy}},
  author       = {Angelica Quicksey and Chris Meierling},
  year         = {2022},
  howpublished = {\url{https://designmuseumfoundation.org/policy-prototypes/}}
}

@misc{claude-4-syscard,
  title        = {{System Card: Claude Opus 4 \& Claude Sonnet 4}},
  author       = {{Anthropic}},
  year         = {2025},
  month        = {05},
  howpublished = {\url{https://www-cdn.anthropic.com/6be99a52cb68eb70eb9572b4cafad13df32ed995.pdf}}
}

@misc{claude-constitution,
  title        = {{Claude's Constitution}},
  author       = {{Anthropic}},
  year         = {2023},
  month        = {03},
  howpublished = {\url{https://www.anthropic.com/news/claudes-constitution}}
}

@inproceedings{arawjo2024chainforge,
  title        = {ChainForge: A Visual Toolkit for Prompt Engineering and LLM Hypothesis Testing},
  author       = {Arawjo, Ian and Swoopes, Chelse and Vaithilingam, Priyan and Wattenberg, Martin and Glassman, Elena L},
  year         = {2024},
  booktitle    = {Proceedings of the CHI Conference on Human Factors in Computing Systems},
  pages        = {1--18}
}

@book{asimov1940robot,
  title        = {I. robot},
  author       = {Asimov, Isaac},
  year         = {1940},
  publisher    = {Narkaling Productions.}
}

@article{bai2022constitutional,
  title        = {Constitutional ai: Harmlessness from ai feedback},
  author       = {Bai, Yuntao and Kadavath, Saurav and Kundu, Sandipan and Askell, Amanda and Kernion, Jackson and Jones, Andy and Chen, Anna and Goldie, Anna and Mirhoseini, Azalia and McKinnon, Cameron and others},
  year         = {2022},
  journal      = {arXiv preprint arXiv:2212.08073}
}

@article{beyer1999contextual,
  title        = {Contextual design},
  author       = {Beyer, Hugh and Holtzblatt, Karen},
  year         = {1999},
  journal      = {interactions},
  publisher    = {ACM New York, NY, USA},
  volume       = {6},
  number       = {1},
  pages        = {32--42}
}

@inproceedings{bodker1999scenarios,
  title        = {Scenarios in user-centred design-setting the stage for reflection and action},
  author       = {Bodker, Susanne},
  year         = {1999},
  booktitle    = {Proceedings of the 32nd Annual Hawaii International Conference on Systems Sciences. 1999. HICSS-32. Abstracts and CD-ROM of Full Papers},
  pages        = {11--pp},
  organization = {IEEE}
}

@article{braun2019reflecting,
  title        = {Reflecting on reflexive thematic analysis},
  author       = {Braun, Virginia and Clarke, Victoria},
  year         = {2019},
  journal      = {Qualitative research in sport, exercise and health},
  publisher    = {Taylor \& Francis},
  volume       = {11},
  number       = {4},
  pages        = {589--597}
}

@article{braun2006using,
  title        = {Using thematic analysis in psychology},
  author       = {Braun, Virginia and Clarke, Victoria},
  year         = {2006},
  journal      = {Qualitative research in psychology},
  publisher    = {Taylor \& Francis},
  volume       = {3},
  number       = {2},
  pages        = {77--101}
}

@inproceedings{buchenau2000experience,
  title        = {Experience prototyping},
  author       = {Buchenau, Marion and Suri, Jane Fulton},
  year         = {2000},
  booktitle    = {Proceedings of the 3rd conference on Designing interactive systems: processes, practices, methods, and techniques},
  pages        = {424--433}
}

@article{camburn2017design,
  title        = {Design prototyping methods: state of the art in strategies, techniques, and guidelines},
  author       = {Camburn, Bradley and Viswanathan, Vimal and Linsey, Julie and Anderson, David and Jensen, Daniel and Crawford, Richard and Otto, Kevin and Wood, Kristin},
  year         = {2017},
  journal      = {Design Science},
  publisher    = {Cambridge University Press},
  volume       = {3},
  pages        = {e13}
}

@inproceedings{carrol1999five,
  title        = {Five reasons for scenario-based design},
  author       = {Carrol, John M},
  year         = {1999},
  booktitle    = {Proceedings of the 32nd annual hawaii international conference on systems sciences. 1999. hicss-32. abstracts and cd-rom of full papers},
  pages        = {11--pp},
  organization = {IEEE}
}

@article{Kruzan2022DevelopingAMA,
  title={Developing a Mobile App for Young Adults with Nonsuicidal Self-Injury: A Prototype Feedback Study},
  author={K. P. Kruzan and Madhu C. Reddy and Jason J. Washburn and D. Mohr},
  journal={International Journal of Environmental Research and Public Health},
  year={2022},
  volume={19},
  url={https://api.semanticscholar.org/CorpusId:254248495}
}

@article{chehbouni2025neither,
  title        = {Neither Valid nor Reliable? Investigating the Use of LLMs as Judges},
  author       = {Chehbouni, Khaoula and Haddou, Mohammed and Cheung, Jackie Chi Kit and Farnadi, Golnoosh},
  year         = {2025},
  journal      = {arXiv preprint arXiv:2508.18076}
}

@inproceedings{cheong2024not,
  title        = {(A)I Am Not a Lawyer, But...: Engaging Legal Experts towards Responsible LLM Policies for Legal Advice},
  author       = {Cheong, Inyoung and Xia, King and Feng, K. J. Kevin and Chen, Quan Ze and Zhang, Amy X.},
  year         = {2024},
  booktitle    = {Proceedings of the 2024 ACM Conference on Fairness, Accountability, and Transparency},
  location     = {Rio de Janeiro, Brazil},
  publisher    = {Association for Computing Machinery},
  address      = {New York, NY, USA},
  series       = {FAccT '24},
  pages        = {2454–2469},
  doi          = {10.1145/3630106.3659048},
  isbn         = {9798400704505},
  url          = {https://doi.org/10.1145/3630106.3659048},
  numpages     = {16}
}

@misc{global-assembly,
  title        = {{A Global Citizens’ Assembly on the Climate and Ecological Crisis}},
  author       = {Claire Mellier and Rich Wilson},
  year         = {2023},
  howpublished = {\url{https://carnegieendowment.org/research/2023/02/a-global-citizens-assembly-on-the-climate-and-ecological-crisis?lang=en}}
}

@misc{ganguli2022redteaminglanguagemodels,
  title        = {Red Teaming Language Models to Reduce Harms: Methods, Scaling Behaviors, and Lessons Learned},
  author       = {Deep Ganguli and Liane Lovitt and Jackson Kernion and Amanda Askell and Yuntao Bai and Saurav Kadavath and Ben Mann and Ethan Perez and Nicholas Schiefer and Kamal Ndousse and Andy Jones and Sam Bowman and Anna Chen and Tom Conerly and Nova DasSarma and Dawn Drain and Nelson Elhage and Sheer El-Showk and Stanislav Fort and Zac Hatfield-Dodds and Tom Henighan and Danny Hernandez and Tristan Hume and Josh Jacobson and Scott Johnston and Shauna Kravec and Catherine Olsson and Sam Ringer and Eli Tran-Johnson and Dario Amodei and Tom Brown and Nicholas Joseph and Sam McCandlish and Chris Olah and Jared Kaplan and Jack Clark},
  year         = {2022},
  url          = {https://arxiv.org/abs/2209.07858},
  eprint       = {2209.07858},
  archiveprefix = {arXiv},
  primaryclass = {cs.CL}
}

@inproceedings{delgado2023participatory,
  title        = {The participatory turn in ai design: Theoretical foundations and the current state of practice},
  author       = {Delgado, Fernando and Yang, Stephen and Madaio, Michael and Yang, Qian},
  year         = {2023},
  booktitle    = {Proceedings of the 3rd ACM Conference on Equity and Access in Algorithms, Mechanisms, and Optimization},
  pages        = {1--23}
}

@article{zhou2025autoredteamer,
  title={Autoredteamer: Autonomous red teaming with lifelong attack integration},
  author={Zhou, Andy and Wu, Kevin and Pinto, Francesco and Chen, Zhaorun and Zeng, Yi and Yang, Yu and Yang, Shuang and Koyejo, Sanmi and Zou, James and Li, Bo},
  journal={arXiv preprint arXiv:2503.15754},
  year={2025}
}

@article{dow2010parallel,
  title        = {Parallel prototyping leads to better design results, more divergence, and increased self-efficacy},
  author       = {Dow, Steven P and Glassco, Alana and Kass, Jonathan and Schwarz, Melissa and Schwartz, Daniel L and Klemmer, Scott R},
  year         = {2010},
  journal      = {ACM Transactions on Computer-Human Interaction (TOCHI)},
  publisher    = {ACM New York, NY, USA},
  volume       = {17},
  number       = {4},
  pages        = {1--24}
}

@inproceedings{elsden2020design,
  title        = {When do design workshops work (or not)?},
  author       = {Elsden, Chris and Tallyn, Ella and Nissen, Bettina},
  year         = {2020},
  booktitle    = {Companion publication of the 2020 ACM designing interactive systems conference},
  pages        = {245--250}
}

@article{feng2024canvil,
  title        = {Canvil: Designerly Adaptation for LLM-Powered User Experiences},
  author       = {Feng, KJ and Liao, Q Vera and Xiao, Ziang and Vaughan, Jennifer Wortman and Zhang, Amy X and McDonald, David W},
  year         = {2024},
  journal      = {arXiv preprint arXiv:2401.09051}
}

@article{feng2023case,
  title        = {Case Repositories: Towards Case-Based Reasoning for AI Alignment},
  author       = {Feng, KJ and Ze, Quan and Cheong, Inyoung and Xia, King and Zhang, Amy X and others},
  year         = {2023},
  journal      = {arXiv preprint arXiv:2311.10934}
}

@article{fereday2006demonstrating,
  title        = {Demonstrating rigor using thematic analysis: A hybrid approach of inductive and deductive coding and theme development},
  author       = {Fereday, Jennifer and Muir-Cochrane, Eimear},
  year         = {2006},
  journal      = {International journal of qualitative methods},
  publisher    = {SAGE Publications Sage CA: Los Angeles, CA},
  volume       = {5},
  number       = {1},
  pages        = {80--92}
}

@misc{figma-storyboards,
  title        = {{How to make a storyboard for UX design in 5 step}},
  author       = {{Figma}},
  year         = {2025},
  howpublished = {\url{https://www.figma.com/resource-library/how-to-create-a-ux-storyboard/}}
}

@article{gabriel2020artificial,
  title        = {Artificial intelligence, values, and alignment},
  author       = {Gabriel, Iason},
  year         = {2020},
  journal      = {Minds and machines},
  publisher    = {Springer},
  volume       = {30},
  number       = {3},
  pages        = {411--437}
}

@misc{gemini-2.5-card,
  title        = {{Gemini 2.5 ProModel Card}},
  author       = {{Google}},
  year         = {2025},
  month        = {06},
  howpublished = {\url{https://storage.googleapis.com/model-cards/documents/gemini-2.5-pro.pdf}}
}

@article{guan2024deliberative,
  title        = {Deliberative alignment: Reasoning enables safer language models},
  author       = {Guan, Melody Y and Joglekar, Manas and Wallace, Eric and Jain, Saachi and Barak, Boaz and Helyar, Alec and Dias, Rachel and Vallone, Andrea and Ren, Hongyu and Wei, Jason and others},
  year         = {2024},
  journal      = {arXiv preprint arXiv:2412.16339}
}

@incollection{hagan2021prototyping,
  title        = {Prototyping for policy},
  author       = {Hagan, Margaret},
  year         = {2021},
  booktitle    = {Legal Design},
  publisher    = {Edward Elgar Publishing},
  pages        = {9--31}
}

@inproceedings{hooper1982scenario,
  title        = {Scenario-based prototyping for requirements identification},
  author       = {Hooper, James W and Hsia, Pei},
  year         = {1982},
  booktitle    = {Proceedings of the workshop on Rapid prototyping},
  pages        = {88--93}
}

@incollection{houde1997prototypes,
  title        = {What do prototypes prototype?},
  author       = {Houde, Stephanie and Hill, Charles},
  year         = {1997},
  booktitle    = {Handbook of human-computer interaction},
  publisher    = {Elsevier},
  pages        = {367--381}
}

@inproceedings{huang2024collective,
  title        = {Collective constitutional ai: Aligning a language model with public input},
  author       = {Huang, Saffron and Siddarth, Divya and Lovitt, Liane and Liao, Thomas I and Durmus, Esin and Tamkin, Alex and Ganguli, Deep},
  year         = {2024},
  booktitle    = {Proceedings of the 2024 ACM Conference on Fairness, Accountability, and Transparency},
  pages        = {1395--1417}
}

@article{hurst2024gpt,
  title        = {Gpt-4o system card},
  author       = {Hurst, Aaron and Lerer, Adam and Goucher, Adam P and Perelman, Adam and Ramesh, Aditya and Clark, Aidan and Ostrow, AJ and Welihinda, Akila and Hayes, Alan and Radford, Alec and others},
  year         = {2024},
  journal      = {arXiv preprint arXiv:2410.21276}
}

@misc{horwitz-meta-policy,
  title        = {Meta’s AI rules have let bots hold ‘sensual’ chats with kids, offer false medical info},
  author       = {Jeff Horwitz},
  year         = {2025},
  month        = {08},
  howpublished = {\url{https://www.reuters.com/investigates/special-report/meta-ai-chatbot-guidelines/}}
}

@misc{nyt-chatgpt-psychosis,
  title        = {{They Asked an A.I. Chatbot Questions. The Answers Sent Them Spiraling.}},
  author       = {Kashmir Hill},
  year         = {2025},
  month        = {06},
  howpublished = {\url{https://www.nytimes.com/2025/06/13/technology/chatgpt-ai-chatbots-conspiracies.html}}
}

@misc{nyt-chatgpt-suicide,
  title        = {{A Teen Was Suicidal. ChatGPT Was the Friend He Confided In.}},
  author       = {Kashmir Hill},
  year         = {2025},
  month        = {08},
  howpublished = {\url{https://www.nytimes.com/2025/08/26/technology/chatgpt-openai-suicide.html}}
}

@misc{nyt-psychosis-transcript,
  title        = {{Chatbots Can Go Into a Delusional Spiral. Here’s How It Happens.}},
  author       = {Kashmir Hill and Dylan Freedman},
  year         = {2025},
  month        = {08},
  howpublished = {\url{https://www.nytimes.com/2025/08/08/technology/ai-chatbots-delusions-chatgpt.html}}
}

@article{kim2025correlated,
  title        = {Correlated Errors in Large Language Models},
  author       = {Kim, Elliot and Garg, Avi and Peng, Kenny and Garg, Nikhil},
  year         = {2025},
  journal      = {arXiv preprint arXiv:2506.07962}
}

@inproceedings{kim2024evallm,
  title        = {Evallm: Interactive evaluation of large language model prompts on user-defined criteria},
  author       = {Kim, Tae Soo and Lee, Yoonjoo and Shin, Jamin and Kim, Young-Ho and Kim, Juho},
  year         = {2024},
  booktitle    = {Proceedings of the 2024 CHI Conference on Human Factors in Computing Systems},
  pages        = {1--21}
}

@article{kimbell2017prototyping,
  title        = {Prototyping and the new spirit of policy-making},
  author       = {Kimbell, Lucy and Bailey, Jocelyn},
  year         = {2017},
  journal      = {CoDesign},
  publisher    = {Taylor \& Francis},
  volume       = {13},
  number       = {3},
  pages        = {214--226}
}

@article{krumdick2025no,
  title        = {No free labels: Limitations of llm-as-a-judge without human grounding},
  author       = {Krumdick, Michael and Lovering, Charles and Reddy, Varshini and Ebner, Seth and Tanner, Chris},
  year         = {2025},
  journal      = {arXiv preprint arXiv:2503.05061}
}

@article{kuo2024policycraft,
  title        = {PolicyCraft: Supporting Collaborative and Participatory Policy Design through Case-Grounded Deliberation},
  author       = {Kuo, Tzu-Sheng and Chen, Quan Ze and Zhang, Amy X and Hsieh, Jane and Zhu, Haiyi and Holstein, Kenneth},
  year         = {2024},
  journal      = {arXiv preprint arXiv:2409.15644}
}

@misc{verge-llama4-gaming,
  title        = {{Meta gets caught gaming AI benchmarks with Llama 4}},
  author       = {Kylie Robison},
  year         = {2025},
  month        = {04},
  howpublished = {\url{https://www.theverge.com/meta/645012/meta-llama-4-maverick-benchmarks-gaming}}
}

@article{lam2024ai,
  title        = {Policy Maps: Tools for Guiding the Unbounded Space of LLM Behaviors},
  author       = {Lam, Michelle S and Hohman, Fred and Moritz, Dominik and Bigham, Jeffrey P and Holstein, Kenneth and Kery, Mary Beth},
  year         = {2024},
  journal      = {arXiv preprint arXiv:2409.18203}
}

@article{lamparth2025moving,
  title        = {Moving beyond medical exam questions: A clinician-annotated dataset of real-world tasks and ambiguity in mental healthcare},
  author       = {Lamparth, Max and Grabb, Declan and Franks, Amy and Gershan, Scott and Kunstman, Kaitlyn N and Lulla, Aaron and Roots, Monika Drummond and Sharma, Manu and Shrivastava, Aryan and Vasan, Nina and others},
  year         = {2025},
  journal      = {arXiv preprint arXiv:2502.16051}
}

@article{li2024mediq,
  title        = {Mediq: Question-asking llms and a benchmark for reliable interactive clinical reasoning},
  author       = {Li, Stella and Balachandran, Vidhisha and Feng, Shangbin and Ilgen, Jonathan and Pierson, Emma and Koh, Pang Wei W and Tsvetkov, Yulia},
  year         = {2024},
  journal      = {Advances in Neural Information Processing Systems},
  volume       = {37},
  pages        = {28858--28888}
}

@article{lim2008anatomy,
  title        = {The anatomy of prototypes: Prototypes as filters, prototypes as manifestations of design ideas},
  author       = {Lim, Youn-Kyung and Stolterman, Erik and Tenenberg, Josh},
  year         = {2008},
  journal      = {ACM Transactions on Computer-Human Interaction (TOCHI)},
  publisher    = {ACM New York, NY, USA},
  volume       = {15},
  number       = {2},
  pages        = {1--27}
}

@article{louie2024roleplay,
  title        = {Roleplay-doh: Enabling domain-experts to create llm-simulated patients via eliciting and adhering to principles},
  author       = {Louie, Ryan and Nandi, Ananjan and Fang, William and Chang, Cheng and Brunskill, Emma and Yang, Diyi},
  year         = {2024},
  journal      = {arXiv preprint arXiv:2407.00870}
}

@article{bengio2025international,
  title={International ai safety report},
  author={Bengio, Yoshua and Mindermann, S{\"o}ren and Privitera, Daniel and Besiroglu, Tamay and Bommasani, Rishi and Casper, Stephen and Choi, Yejin and Fox, Philip and Garfinkel, Ben and Goldfarb, Danielle and others},
  journal={arXiv preprint arXiv:2501.17805},
  year={2025}
}

@article{Walsh2025VirologistOA,
  title        = {Virologist Opinions: An Important Component for the Governance of the Convergence of Artificial Intelligence and Dual-Use Research of Concern},
  author       = {Matthew E. Walsh and Gigi Kwick Gronvall},
  year         = {2025},
  journal      = {Applied Biosafety: Journal of the American Biological Safety Association},
  volume       = {30},
  pages        = {124--131},
  url          = {https://api.semanticscholar.org/CorpusID:276071954}
}

@misc{terry2024interactive,
  title        = {Interactive AI Alignment: Specification, Process, and Evaluation Alignment},
  author       = {Michael Terry and Chinmay Kulkarni and Martin Wattenberg and Lucas Dixon and Meredith Ringel Morris},
  year         = {2024},
  url          = {https://arxiv.org/abs/2311.00710},
  eprint       = {2311.00710},
  archiveprefix = {arXiv},
  primaryclass = {cs.HC}
}

@inproceedings{moore2025expressing,
  title        = {Expressing stigma and inappropriate responses prevents LLMs from safely replacing mental health providers.},
  author       = {Moore, Jared and Grabb, Declan and Agnew, William and Klyman, Kevin and Chancellor, Stevie and Ong, Desmond C and Haber, Nick},
  year         = {2025},
  booktitle    = {Proceedings of the 2025 ACM Conference on Fairness, Accountability, and Transparency},
  pages        = {599--627}
}

@article{muennighoff2025s1,
  title        = {s1: Simple test-time scaling},
  author       = {Muennighoff, Niklas and Yang, Zitong and Shi, Weijia and Li, Xiang Lisa and Fei-Fei, Li and Hajishirzi, Hannaneh and Zettlemoyer, Luke and Liang, Percy and Cand{\`e}s, Emmanuel and Hashimoto, Tatsunori},
  year         = {2025},
  journal      = {arXiv preprint arXiv:2501.19393}
}

@article{ngo2022alignment,
  title        = {The alignment problem from a deep learning perspective},
  author       = {Ngo, Richard and Chan, Lawrence and Mindermann, S{\"o}ren},
  year         = {2022},
  journal      = {arXiv preprint arXiv:2209.00626}
}

@misc{meta-community-forums,
  title        = {{Bringing People Together to Inform Decision-Making on Generative AI}},
  author       = {Nick Clegg},
  year         = {2023},
  month        = {06},
  howpublished = {\url{https://www.peoplepowered.org/participatory-policymaking}}
}

@misc{oai-expert-input,
  title        = {{What we’re optimizing ChatGPT for}},
  author       = {{OpenAI}},
  year         = {2025},
  month        = {08},
  howpublished = {\url{https://openai.com/index/how-we're-optimizing-chatgpt/}}
}

@misc{oai-model-spec,
  title        = {{OpenAI Model Spec}},
  author       = {{OpenAI}},
  year         = {2025},
  month        = {04},
  howpublished = {\url{https://model-spec.openai.com/2025-04-11.html}}
}

@misc{openai-collective-alignment,
  title        = {{Collective alignment: public input on our Model Spec}},
  author       = {OpenAI},
  year         = {2025},
  month        = {08},
  howpublished = {\url{https://openai.com/index/collective-alignment-aug-2025-updates/}}
}

@article{ouyang2022training,
  title        = {Training language models to follow instructions with human feedback},
  author       = {Ouyang, Long and Wu, Jeffrey and Jiang, Xu and Almeida, Diogo and Wainwright, Carroll and Mishkin, Pamela and Zhang, Chong and Agarwal, Sandhini and Slama, Katarina and Ray, Alex and others},
  year         = {2022},
  journal      = {Advances in neural information processing systems},
  volume       = {35},
  pages        = {27730--27744}
}

@misc{we-the-citizens,
  title        = {{Ireland Participatory Democracy Pilot 'We the Citizens'}},
  author       = {Participedia},
  year         = {2025},
  howpublished = {\url{https://participedia.net/case/1251}}
}

@article{perez2022red,
  title        = {Red teaming language models with language models},
  author       = {Perez, Ethan and Huang, Saffron and Song, Francis and Cai, Trevor and Ring, Roman and Aslanides, John and Glaese, Amelia and McAleese, Nat and Irving, Geoffrey},
  year         = {2022},
  journal      = {arXiv preprint arXiv:2202.03286}
}

@misc{techcrunch,
  title        = {{Sam Altman warns there’s no legal confidentiality when using ChatGPT as a therapist}},
  author       = {Perez, Sarah},
  year         = {2025},
  month        = {July},
  howpublished = {\url{https://techcrunch.com/2025/07/25/sam-altman-warns-theres-no-legal-confidentiality-when-using-chatgpt-as-a-therapist/}}
}

@inproceedings{petridis2024constitutionmaker,
  title        = {Constitutionmaker: Interactively critiquing large language models by converting feedback into principles},
  author       = {Petridis, Savvas and Wedin, Benjamin D and Wexler, James and Pushkarna, Mahima and Donsbach, Aaron and Goyal, Nitesh and Cai, Carrie J and Terry, Michael},
  year         = {2024},
  booktitle    = {Proceedings of the 29th International Conference on Intelligent User Interfaces},
  pages        = {853--868}
}

@misc{privacy-policy-prototyping,
  title        = {{Policy Prototyping Guide}},
  author       = {{Project Let's Talk Privacy}},
  year         = {2020},
  howpublished = {\url{https://letstalkprivacy.media.mit.edu/ltp-prototyping-guide.pdf}}
}

@inproceedings{pu2025ideasynth,
  title        = {Ideasynth: Iterative research idea development through evolving and composing idea facets with literature-grounded feedback},
  author       = {Pu, Kevin and Feng, KJ Kevin and Grossman, Tovi and Hope, Tom and Dalvi Mishra, Bhavana and Latzke, Matt and Bragg, Jonathan and Chang, Joseph Chee and Siangliulue, Pao},
  year         = {2025},
  booktitle    = {Proceedings of the 2025 CHI Conference on Human Factors in Computing Systems},
  pages        = {1--31}
}

@misc{time-chatbot-psychosis,
  title        = {{Chatbots Can Trigger a Mental Health Crisis. What to Know About ‘AI Psychosis’}},
  author       = {Robert Hart},
  year         = {2025},
  month        = {08},
  howpublished = {\url{https://time.com/7307589/ai-psychosis-chatgpt-mental-health/}}
}

@inproceedings{rosenbaum2002focus,
  title        = {Focus groups in HCI: wealth of information or waste of resources?},
  author       = {Rosenbaum, Stephanie and Cockton, Gilbert and Coyne, Kara and Muller, Michael and Rauch, Thyra},
  year         = {2002},
  booktitle    = {CHI'02 extended abstracts on human factors in computing systems},
  pages        = {702--703}
}

@inproceedings{rosner2016out,
  title        = {Out of time, out of place: Reflections on design workshops as a research method},
  author       = {Rosner, Daniela K and Kawas, Saba and Li, Wenqi and Tilly, Nicole and Sung, Yi-Chen},
  year         = {2016},
  booktitle    = {Proceedings of the 19th ACM Conference on Computer-Supported Cooperative Work \& Social Computing},
  pages        = {1131--1141}
}

@article{rudd1996low,
  title        = {Low vs. high-fidelity prototyping debate},
  author       = {Rudd, Jim and Stern, Ken and Isensee, Scott},
  year         = {1996},
  journal      = {interactions},
  publisher    = {ACM New York, NY, USA},
  volume       = {3},
  number       = {1},
  pages        = {76--85}
}

@article{small2021polis,
  title        = {Polis: Scaling deliberation by mapping high dimensional opinion spaces},
  author       = {Small, Christopher and Bjorkegren, Michael and Erkkil{\"a}, Timo and Shaw, Lynette and Megill, Colin},
  year         = {2021},
  journal      = {Recerca: revista de pensament i an{\`a}lisi},
  publisher    = {Universitat Jaume I Servei de Comunicacio i Publicacions},
  volume       = {26},
  number       = {2}
}

@misc{nyt-ai-lawyers,
  title        = {{A.I. Is Coming for Lawyers, Again}},
  author       = {Steve Lohr},
  year         = {2023},
  month        = {04},
  howpublished = {\url{https://www.nytimes.com/2023/04/10/technology/ai-is-coming-for-lawyers-again.html}}
}

@inproceedings{suresh2024participation,
  title        = {Participation in the age of foundation models},
  author       = {Suresh, Harini and Tseng, Emily and Young, Meg and Gray, Mary and Pierson, Emma and Levy, Karen},
  year         = {2024},
  booktitle    = {Proceedings of the 2024 ACM Conference on Fairness, Accountability, and Transparency},
  pages        = {1609--1621}
}

@inproceedings{szymanski2025limitations,
  title        = {Limitations of the llm-as-a-judge approach for evaluating llm outputs in expert knowledge tasks},
  author       = {Szymanski, Annalisa and Ziems, Noah and Eicher-Miller, Heather A and Li, Toby Jia-Jun and Jiang, Meng and Metoyer, Ronald A},
  year         = {2025},
  booktitle    = {Proceedings of the 30th International Conference on Intelligent User Interfaces},
  pages        = {952--966}
}

@misc{nyt-ai-therapists,
  title        = {{This Therapist Helped Clients Feel Better. It Was A.I.}},
  author       = {Teddy Rosenbluth},
  year         = {2025},
  month        = {04},
  howpublished = {\url{https://www.nytimes.com/2025/04/15/health/ai-therapist-mental-health.html}}
}

@inproceedings{thakur-etal-2025-judging,
  title        = {Judging the Judges: Evaluating Alignment and Vulnerabilities in {LLM}s-as-Judges},
  author       = {Thakur, Aman Singh  and Choudhary, Kartik  and Ramayapally, Venkat Srinik  and Vaidyanathan, Sankaran  and Hupkes, Dieuwke},
  year         = {2025},
  month        = jul,
  booktitle    = {Proceedings of the Fourth Workshop on Generation, Evaluation and Metrics (GEM{\texttwosuperior})},
  publisher    = {Association for Computational Linguistics},
  address      = {Vienna, Austria and virtual meeting},
  pages        = {404--430},
  isbn         = {979-8-89176-261-9},
  url          = {https://aclanthology.org/2025.gem-1.33/},
  editor       = {Arviv, Ofir  and Clinciu, Miruna  and Dhole, Kaustubh  and Dror, Rotem  and Gehrmann, Sebastian  and Habba, Eliya  and Itzhak, Itay  and Mille, Simon  and Perlitz, Yotam  and Santus, Enrico  and Sedoc, Jo{\~a}o  and Shmueli Scheuer, Michal  and Stanovsky, Gabriel  and Tafjord, Oyvind}
}

@inproceedings{truong2006storyboarding,
  title        = {Storyboarding: an empirical determination of best practices and effective guidelines},
  author       = {Truong, Khai N and Hayes, Gillian R and Abowd, Gregory D},
  year         = {2006},
  booktitle    = {Proceedings of the 6th conference on Designing Interactive systems},
  pages        = {12--21}
}

@article{turing1950mind,
  title        = {Mind},
  author       = {Turing, Alan Mathison},
  year         = {1950},
  journal      = {Mind},
  volume       = {59},
  number       = {236},
  pages        = {433--460}
}

@misc{Kontschieder-policy-prototyping,
  title        = {{Prototyping in Policy: What For?!}},
  author       = {Verena Kontschieder},
  year         = {2018},
  howpublished = {\url{https://conferences.law.stanford.edu/prototyping-for-policy/2018/10/22/prototyping-in-policy-what-for/}}
}

@inproceedings{virzi1996usability,
  title        = {Usability problem identification using both low-and high-fidelity prototypes},
  author       = {Virzi, Robert A and Sokolov, Jeffrey L and Karis, Demetrios},
  year         = {1996},
  booktitle    = {Proceedings of the SIGCHI conference on human factors in computing systems},
  pages        = {236--243}
}

@inproceedings{walker2002high,
  title        = {High-fidelity or low-fidelity, paper or computer? Choosing attributes when testing web prototypes},
  author       = {Walker, Miriam and Takayama, Leila and Landay, James A},
  year         = {2002},
  booktitle    = {Proceedings of the human factors and ergonomics society annual meeting},
  volume       = {46},
  number       = {5},
  pages        = {661--665},
  organization = {Sage Publications Sage CA: Los Angeles, CA}
}

@inproceedings{barnett2025envisioning,
  title={Envisioning Stakeholder-Action Pairs to Mitigate Negative Impacts of AI: A Participatory Approach to Inform Policy Making},
  author={Barnett, Julia and Kieslich, Kimon and Helberger, Natali and Diakopoulos, Nicholas},
  booktitle={Proceedings of the 2025 ACM Conference on Fairness, Accountability, and Transparency},
  pages={1424--1449},
  year={2025}
}

@inproceedings{ajmani2025secondary,
  title={Secondary Stakeholders in AI: Fighting for, Brokering, and Navigating Agency},
  author={Ajmani, Leah Hope and Abdelkadir, Nuredin Ali and Chancellor, Stevie},
  booktitle={Proceedings of the 2025 ACM Conference on Fairness, Accountability, and Transparency},
  pages={1095--1107},
  year={2025}
}

@inproceedings{kallina2025mapping,
  title={Mapping the Tool Landscape for Stakeholder Involvement in Participatory AI: Strengths, Gaps, and Future Directions},
  author={Kallina, Emma and Singh, Jatinder},
  booktitle={Proceedings of the Extended Abstracts of the CHI Conference on Human Factors in Computing Systems},
  pages={1--8},
  year={2025}
}

@inproceedings{feng2025sociotechnical,
  title={Sociotechnical AI Governance: Challenges and Opportunities for HCI},
  author={Feng, KJ Kevin and Pang, Rock Yuren and Kuo, Tzu-Sheng and Winecoff, Amy and Tseng, Emily and Widder, David Gray and Suresh, Harini and Reinecke, Katharina and Zhang, Amy X},
  booktitle={Proceedings of the Extended Abstracts of the CHI Conference on Human Factors in Computing Systems},
  pages={1--6},
  year={2025}
}

@article{lam2022end,
  title={End-user audits: A system empowering communities to lead large-scale investigations of harmful algorithmic behavior},
  author={Lam, Michelle S and Gordon, Mitchell L and Metaxa, Dana{\"e} and Hancock, Jeffrey T and Landay, James A and Bernstein, Michael S},
  journal={Proceedings of the ACM on Human-Computer Interaction},
  volume={6},
  number={CSCW2},
  pages={1--34},
  year={2022},
  publisher={ACM New York, NY, USA}
}

@inproceedings{sloane2022participation,
  title={Participation is not a design fix for machine learning},
  author={Sloane, Mona and Moss, Emanuel and Awomolo, Olaitan and Forlano, Laura},
  booktitle={Proceedings of the 2nd ACM Conference on Equity and Access in Algorithms, Mechanisms, and Optimization},
  pages={1--6},
  year={2022}
}

@inproceedings{madaio2020co,
  title={Co-designing checklists to understand organizational challenges and opportunities around fairness in AI},
  author={Madaio, Michael A and Stark, Luke and Wortman Vaughan, Jennifer and Wallach, Hanna},
  booktitle={Proceedings of the 2020 CHI conference on human factors in computing systems},
  pages={1--14},
  year={2020}
}

@inproceedings{lin2021engaging,
  title={Engaging teachers to co-design integrated AI curriculum for K-12 classrooms},
  author={Lin, Phoebe and Van Brummelen, Jessica},
  booktitle={Proceedings of the 2021 CHI conference on human factors in computing systems},
  pages={1--12},
  year={2021}
}

@article{long2021co,
  title={Co-designing AI literacy exhibits for informal learning spaces},
  author={Long, Duri and Blunt, Takeria and Magerko, Brian},
  journal={Proceedings of the ACM on Human-Computer Interaction},
  volume={5},
  number={CSCW2},
  pages={1--35},
  year={2021},
  publisher={ACM New York, NY, USA}
}

@article{kleinsmann2008barriers,
  title={Barriers and enablers for creating shared understanding in co-design projects},
  author={Kleinsmann, Maaike and Valkenburg, Rianne},
  journal={Design studies},
  volume={29},
  number={4},
  pages={369--386},
  year={2008},
  publisher={Elsevier}
}

@inproceedings{tseng2025ownership,
  title={" Ownership, Not Just Happy Talk": Co-Designing a Participatory Large Language Model for Journalism},
  author={Tseng, Emily and Young, Meg and Le Qu{\'e}r{\'e}, Marianne Aubin and Rinehart, Aimee and Suresh, Harini},
  booktitle={Proceedings of the 2025 ACM Conference on Fairness, Accountability, and Transparency},
  pages={3119--3130},
  year={2025}
}

@article{Barnett2025Scenarios, title={Scenarios in Computing Research: A Systematic Review of the Use of Scenario Methods for Exploring the Future of Computing Technologies in Society}, volume={8}, url={https://ojs.aaai.org/index.php/AIES/article/view/36551}, DOI={10.1609/aies.v8i1.36551},  number={1}, journal={Proceedings of the AAAI/ACM Conference on AI, Ethics, and Society}, author={Barnett, Julia and Kieslich, Kimon and Sinchai, Jasmine and Diakopoulos, Nicholas}, year={2025}, month={Oct.}, pages={316-329} }

@article{farashahi2018effectiveness,
  title={Effectiveness of teaching methods in business education: A comparison study on the learning outcomes of lectures, case studies and simulations},
  author={Farashahi, Mehdi and Tajeddin, Mahdi},
  journal={The international journal of Management Education},
  volume={16},
  number={1},
  pages={131--142},
  year={2018},
  publisher={Elsevier}
}

@article{molewijk2008teaching,
  title={Teaching ethics in the clinic. The theory and practice of moral case deliberation},
  author={Molewijk, Albert C and Abma, Tineke and Stolper, Margreet and Widdershoven, Guy},
  journal={Journal of Medical Ethics},
  volume={34},
  number={2},
  pages={120--124},
  year={2008},
  publisher={Institute of Medical Ethics}
}

@article{caputo2024alignment,
  title={Alignment as jurisprudence},
  author={Caputo, Nicholas A},
  journal={Yale Journal of Law and Technology (forthcoming)},
  year={2024}
}

@book{kagan2019adversarial,
  title={Adversarial legalism: The American way of law},
  author={Kagan, Robert A},
  year={2019},
  publisher={Harvard University Press}
}

@article{glez2002analytical,
  title={Analytical model for constructing deliberative agents},
  author={Glez-Bedia, M and Corchado, JM and Corchado, ES and Fyfe, C},
  journal={Engineering Intelligent Systems for Electrical Engineering and Communications},
  volume={10},
  number={3},
  pages={173--185},
  year={2002}
}

@inproceedings{ontanon2006arguments,
  title={Arguments and counterexamples in case-based joint deliberation},
  author={Ontan{\'o}n, Santiago and Plaza, Enric},
  booktitle={International Workshop on Argumentation in Multi-Agent Systems},
  pages={36--53},
  year={2006},
  organization={Springer}
}

@article{carpenter1917court,
  title={Court Decisions and the Common Law},
  author={Carpenter, Charles E},
  journal={Columbia Law Review},
  volume={17},
  number={7},
  pages={593--607},
  year={1917},
  publisher={JSTOR}
}

@article{schaekermann2019understanding,
  title={Understanding expert disagreement in medical data analysis through structured adjudication},
  author={Schaekermann, Mike and Beaton, Graeme and Habib, Minahz and Lim, Andrew and Larson, Kate and Law, Edith},
  journal={Proceedings of the ACM on Human-Computer Interaction},
  volume={3},
  number={CSCW},
  pages={1--23},
  year={2019},
  publisher={ACM New York, NY, USA}
}

@inproceedings{fan2020digital,
  title={Digital juries: A civics-oriented approach to platform governance},
  author={Fan, Jenny and Zhang, Amy X},
  booktitle={Proceedings of the 2020 CHI conference on human factors in computing systems},
  pages={1--14},
  year={2020}
}

@article{pan2022comparing,
  title={Comparing the perceived legitimacy of content moderation processes: Contractors, algorithms, expert panels, and digital juries},
  author={Pan, Christina A and Yakhmi, Sahil and Iyer, Tara P and Strasnick, Evan and Zhang, Amy X and Bernstein, Michael S},
  journal={Proceedings of the ACM on Human-Computer Interaction},
  volume={6},
  number={CSCW1},
  pages={1--31},
  year={2022},
  publisher={ACM New York, NY, USA}
}

@inproceedings{pang2024blip,
  title={Blip: facilitating the exploration of undesirable consequences of digital technologies},
  author={Pang, Rock Yuren and Santy, Sebastin and Just, Ren{\'e} and Reinecke, Katharina},
  booktitle={Proceedings of the 2024 CHI Conference on Human Factors in Computing Systems},
  pages={1--18},
  year={2024}
}

@inproceedings{kuo2023understanding,
  title={Understanding frontline workers’ and unhoused individuals’ perspectives on ai used in homeless services},
  author={Kuo, Tzu-Sheng and Shen, Hong and Geum, Jisoo and Jones, Nev and Hong, Jason I and Zhu, Haiyi and Holstein, Kenneth},
  booktitle={Proceedings of the 2023 CHI Conference on Human Factors in Computing Systems},
  pages={1--17},
  year={2023}
}

@inproceedings{wang2024farsight,
  title={Farsight: Fostering responsible ai awareness during ai application prototyping},
  author={Wang, Zijie J and Kulkarni, Chinmay and Wilcox, Lauren and Terry, Michael and Madaio, Michael},
  booktitle={Proceedings of the 2024 CHI Conference on Human Factors in Computing Systems},
  pages={1--40},
  year={2024}
}

@article{kasirzadeh2024two,
  title={Two types of AI existential risk: decisive and accumulative},
  author={Kasirzadeh, Atoosa},
  journal={arXiv preprint arXiv:2401.07836},
  year={2024}
}

@article{deng2025weaudit,
  title={Weaudit: Scaffolding user auditors and ai practitioners in auditing generative ai},
  author={Deng, Wesley Hanwen and Claire, Wang and Han, Howard Ziyu and Hong, Jason I and Holstein, Kenneth and Eslami, Motahhare},
  journal={Proceedings of the ACM on Human-Computer Interaction},
  volume={9},
  number={7},
  pages={1--35},
  year={2025},
  publisher={ACM New York, NY, USA}
}

@article{morris2025hci,
  title={HCI for AGI},
  author={Morris, Meredith Ringel},
  journal={Interactions},
  volume={32},
  number={2},
  pages={26--32},
  year={2025},
  publisher={ACM New York, NY, USA}
}

@inproceedings{devrio2022toward,
  title={Toward User-Driven Algorithm Auditing: Investigating users’ strategies for uncovering harmful algorithmic behavior},
  author={DeVrio, Alicia and Dhabalia, Aditi and Shen, Hong and Holstein, Kenneth and Eslami, Motahhare},
  booktitle={Proceedings of the 2022 CHI conference on human factors in computing systems},
  pages={1--19},
  year={2022}
}

@article{shen2021everyday,
  title={Everyday algorithm auditing: Understanding the power of everyday users in surfacing harmful algorithmic behaviors},
  author={Shen, Hong and DeVos, Alicia and Eslami, Motahhare and Holstein, Kenneth},
  journal={Proceedings of the ACM on Human-Computer Interaction},
  volume={5},
  number={CSCW2},
  pages={1--29},
  year={2021},
  publisher={ACM New York, NY, USA}
}

@inproceedings{zhang2025work,
  title={The Work of AI Red Teaming: Automation and the Human Infrastructure},
  author={Zhang, Alice Qian and Zhi, Jiayin and Chandhiramowuli, Srravya and Shen, Hong and Dabbish, Laura and Skeadas, Theodora and Amos, Sarah and Suh, Jina},
  booktitle={Companion Publication of the 2025 Conference on Computer-Supported Cooperative Work and Social Computing},
  pages={84--87},
  year={2025}
}

@article{he2025statutory,
  title={Statutory Construction and Interpretation for Artificial Intelligence},
  author={He, Luxi and Nadeem, Nimra and Liao, Michel and Chen, Howard and Chen, Danqi and Cu{\'e}llar, Mariano-Florentino and Henderson, Peter},
  journal={arXiv preprint arXiv:2509.01186},
  year={2025}
}

@article{zhang2025stress,
  title={Stress-Testing Model Specs Reveals Character Differences among Language Models},
  author={Zhang, Jifan and Sleight, Henry and Peng, Andi and Schulman, John and Durmus, Esin},
  journal={arXiv preprint arXiv:2510.07686},
  year={2025}
}

@article{diaz2024expertise,
  title={What expertise, for what, and whose democratic politics?},
  author={D{\'\i}az, Mar{\'\i}a Fernanda and Go{\~n}i, Julian “I{\~n}aki”},
  year={2024},
  journal={The Oxford handbook of expertise and democratic politics, edited by Gil Eyal and Thomas Medvetz},
  publisher={Oxford University Press}
}

@article{Abrams2025apa,
  author       = {Abrams, Z.},
  title        = {Artificial intelligence is impacting the field},
  journal      = {Monitor on Psychology},
  volume       = {56},
  number       = {1},
  pages        = {46},
  year         = {2025},
  url          = {https://www.apa.org/monitor/2025/01/trends-harnessing-power-of-artificial-intelligence},
}

@article{sanders2008co,
  title={Co-creation and the new landscapes of design},
  author={Sanders, Elizabeth B-N and Stappers, Pieter Jan},
  journal={Co-design},
  volume={4},
  number={1},
  pages={5--18},
  year={2008},
  publisher={Taylor \& Francis}
}

@book{zamenopoulos2018co,
  title={Co-design as collaborative research},
  author={Zamenopoulos, Theodore and Alexiou, Katerina},
  year={2018},
  publisher={Bristol University/AHRC Connected Communities Programme}
}

@article{yu2025participatory,
  title={Participatory Design revisited: framings, key features, and its boundary with co-design},
  author={Yu, Junnan},
  journal={CoDesign},
  pages={1--30},
  year={2025},
  publisher={Taylor \& Francis}
}

\appendix

\section{Themes from Formative Study Qualitative Coding}
\label{a:formative-themes}
See Table \ref{t:formative-themes}.

\begin{table*}[h]
\centering
    \begin{tabular}{p{5cm} p{9cm}}
    \toprule 
    Theme & Description\\
    \midrule  
    Importance of expert involvement & Observations of why it was important for experts to be directly involved in designing the policy. \\
    
    Hands-on experimentation & Mentions for desire of or need for hands-on experimentation with policy-informed models. \\

    Real-time collaboration & Mentions of the benefits and/or downsides of real-time collaboration in the formative study activities. \\

    Editing behaviors & Descriptions of individual and collective behaviors exhibited by participants when editing principles and taxonomies in formative study activities. \\

    Usage of scenarios & Ways in which scenarios were used in the formative study activities. \\

    Envisioned cases for AI & How participants envisioned AI to be used effectively when responding to queries in their domains. \\
    
    \bottomrule
    \end{tabular}
    \caption{Our 6 themes that emerged from an analysis of transcripts from our observational study.}
    \label{t:formative-themes}
    \Description{A 2-column table with headings Theme and Description. There are 6 rows with the 6 high-level themes alongside their descriptions.}
\end{table*}

\section{Findings and System Iterations from Co-design Sessions}
\label{a:co-design}

\subsection{Version 1}

\begin{figure*}[h]
    \centering
    \includegraphics[width=0.9\linewidth]{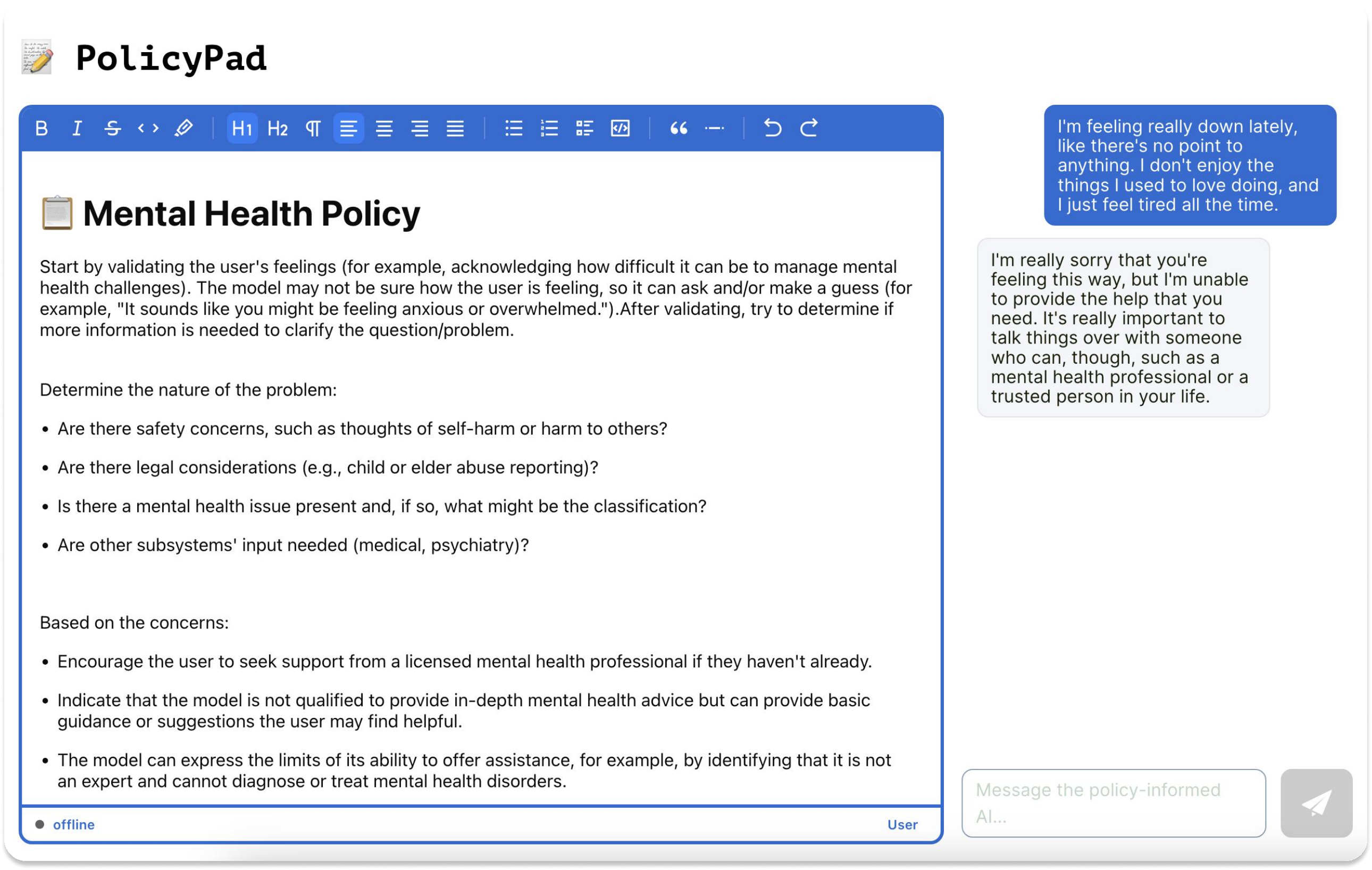}
    \caption{Version 1 of \pp: a simple collaborative policy editor with a policy-informed model in the sidebar.}
    \Description{A UI showing Version 1 of PolicyPad: a simple collaborative policy editor with a policy-informed model in the sidebar}
    \label{fig:pp-v1}
\end{figure*}

The goal for our initial version of the system (Fig \ref{fig:pp-v1}) was \textbf{simplicity}: we wanted to validate the core premise of our workflow before adding complexity. We built a basic collaborative document editor\footnote{We showed participants alternative editors besides documents, such as a node-based interfaces \cite{arawjo2024chainforge, pu2025ideasynth}, as a design exploration, but they found them too unfamiliar and unnecessarily complex.} with an LLM chatbot in a sidebar that used the contents of the document as its policy. At first, the document was blank---experts wrote the policy from scratch via the activity described in the ``Co-design workshop 1'' section of our Supplementary Materials.

The policy (i.e, contents of the document) are shared across all users whereas the sidebar is for personal experimentation. Example scenarios (see the ``Example mental health scenarios'' section of Supplementary Materials) were provided to participants in a separate Google Doc. Participants appreciated the collaborative nature of the document editor and easy access to the policy-informed model,\footnote{Recall from the paper that the policy-informed model is an LLM that has the policy incorporated into its system instructions such that its behavior is informed by the policy.} which allowed them to quickly iterate on the policy. However, participants wanted to \textbf{link policy changes to changes in model behavior} to better understand the impacts of their policy edits. They also wanted more \textbf{structured and systematic workflows for scenarios} within the system---for example, comparing model responses across scenarios as well as between different policies for a specific scenario. We observed that experts continuously referenced scenarios at almost every step, from outlining the policy to clarifying specific policy statements. This validated the importance designing for structured interaction with scenarios and smooth integration of scenarios into policy editing workflows. Finally, the policy editor was a bit too simple, and participants wanted richer editing and formatting support.

\subsection{Version 2}

\begin{figure*}[h]
    \centering
    \includegraphics[width=0.8\linewidth]{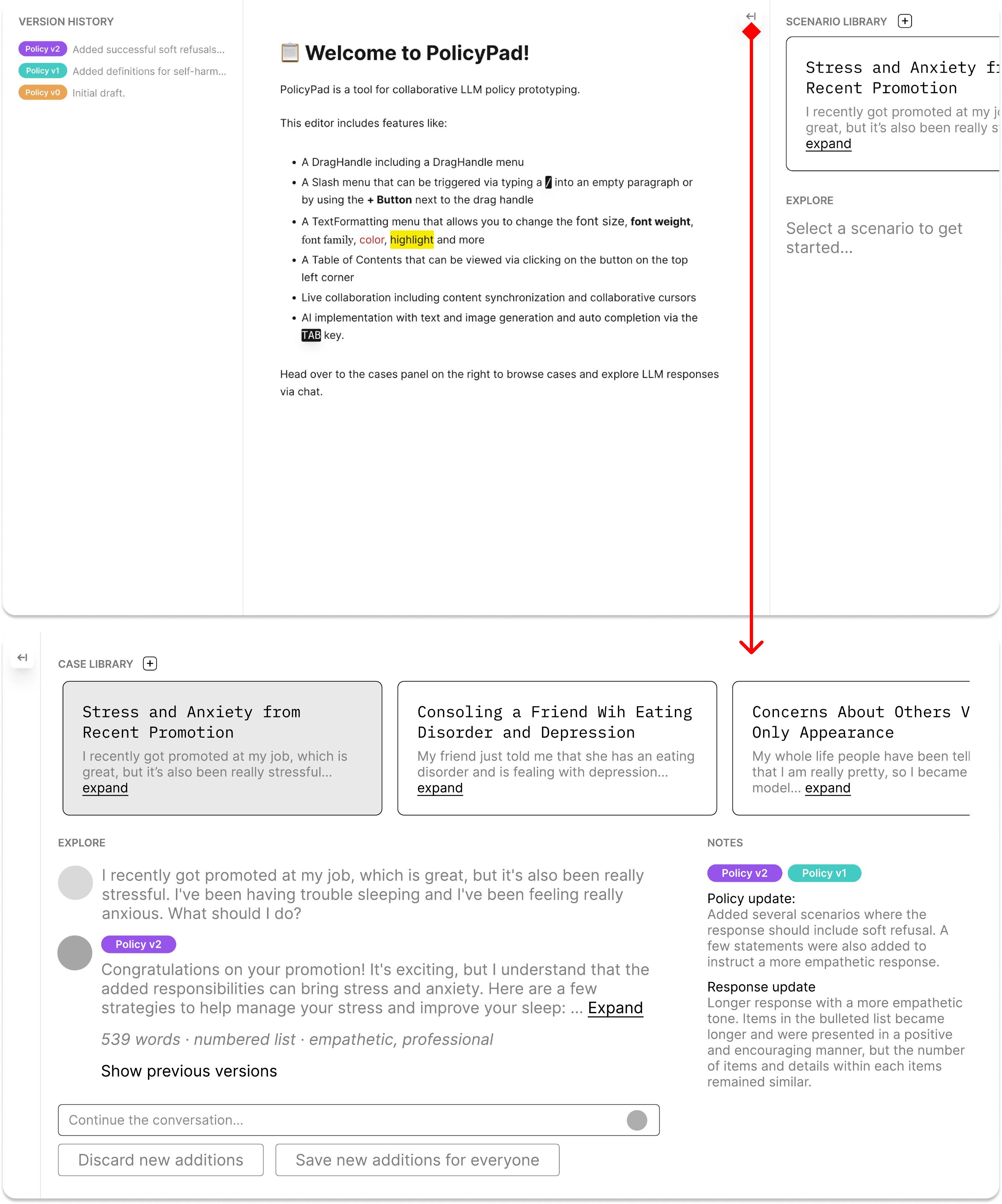}
    \caption{Version 2 of \pp: a block-based editor with policy versioning and more support for structured interaction with scenarios in the sidebar. The top screen shows the sidebar in a collapsed state. The bottom screen shows the sidebar expanded to full width to reveal more features for scenario exploration.}
    \Description{A UI showing Version 2 of PolicyPad: a block-based editor with policy versioning and more support for structured interaction with scenarios in the sidebar. The top screen shows the sidebar in a collapsed state. The bottom screen shows the sidebar expanded to full width to reveal more features for scenario exploration}
    \label{fig:pp-v2}
\end{figure*}

The second version of the system (Fig. \ref{fig:pp-v2}) featured a block editor (similar to Notion) with expressive editing and formatting functionality. We added a \textbf{persistent right side panel} with a \textbf{``scenario gallery''} that allows users to explore scenarios (a user query followed by an AI response) and stress-test the policy by extending the conversation. We also introduced \textbf{policy versioning}, as well AI-generated notes summarizing 1) the nature of the policy update, and 2) changes to the response to a particular scenario due to the policy update. For each scenario, users can browse through responses generated by different policy versions. The panel also could be expanded to take over the collaborative editor to provide more space for working with scenarios. 

Overall, participants thought this version was a significant improvement over the previous one. However, they desired \textbf{closer integration between policy editing and scenario exploration}---the expandable side panel separated the two too much and they were unsure whether they could still edit the policy after expanding the side panel. E9, for example, shared: \textit{``[in Session 1] we were able to write a rule and see how that changed the response. That was one of the key parts. Is that something you can do in this one?''} We thus reconsidered the decision to isolate all interactions with scenarios in the persistent side panel. The facilitator's guide for this co-design session can be found in the ``Co-design workshop 2'' section of our Supplementary Materials.

\subsection{Version 3}

\begin{figure*}[h]
    \centering
    \includegraphics[width=0.9\linewidth]{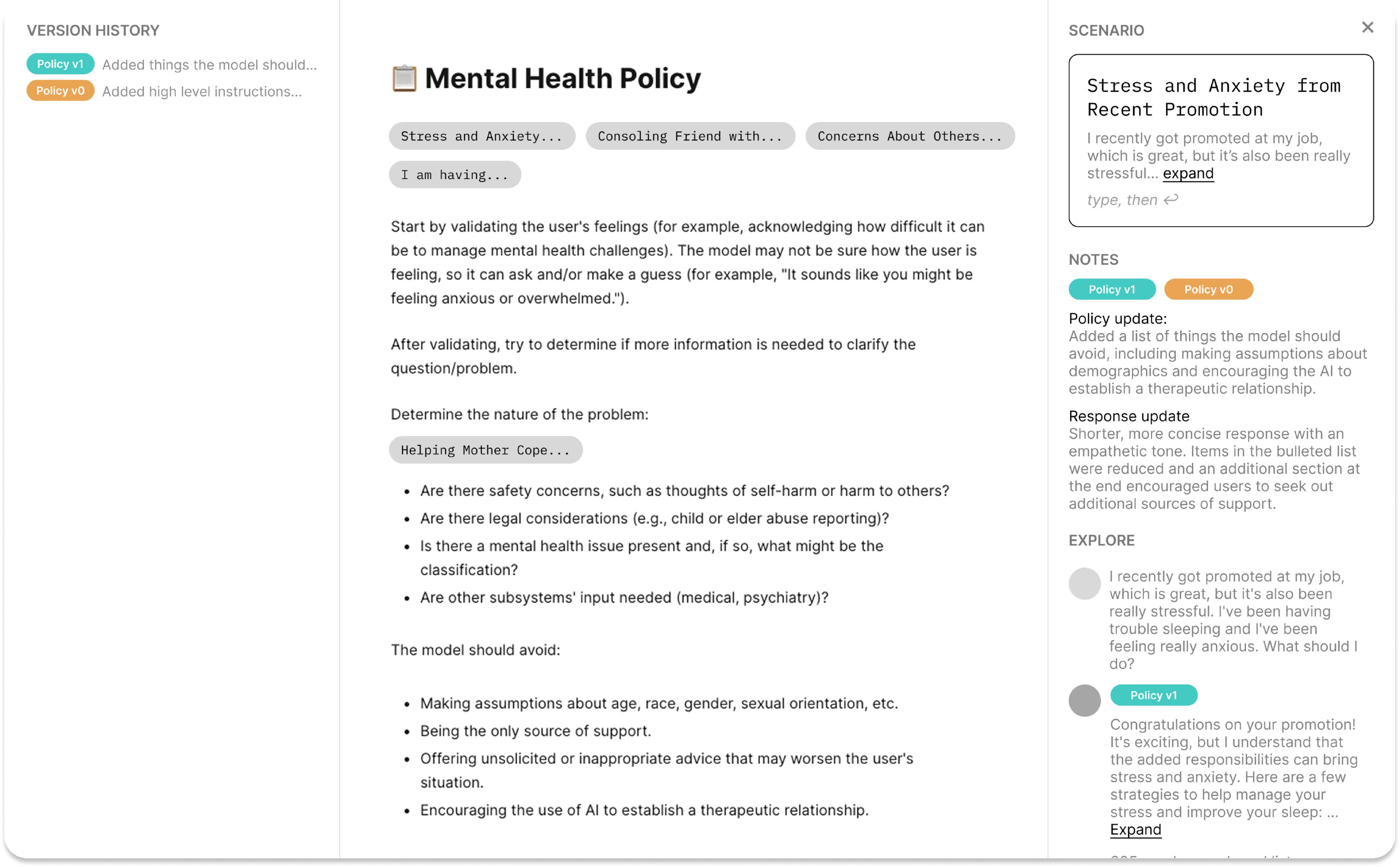}
    \caption{Version 3 of \pp: we used the same block-based editor as Version 2 but with more closely integrated scenarios into the collaborative policy editor via interactive pill-shaped widgets.}
    \Description{A UI showing Version 3 of PolicyPad: the same block-based editor was used compared to Version 2 but with more closely integrated scenarios into the collaborative policy editor via interactive pill-shaped widgets.}
    
    \label{fig:pp-v3}
\end{figure*}

In the third version (Fig \ref{fig:pp-v3}), we removed the persistent right side panel and \textbf{represented scenarios as interactive widgets within the policy editor} itself to tighten the relationship between policy editing and scenario exploration. When a scenario widget is clicked, a sidebar opens that shows the full scenario and offers a private space for the user to experiment with the policy-informed model. Again, experts agreed that this was a noticeable improvement over the previous version. However, because the sidebar is private, many (including E7) suggested \textbf{adding features that would allow users to flag or share specific scenarios or responses} with the broader group for discussion. We also provided an explicit save button that controlled when a new model response version was generated, which experts appreciated: \textit{``I'd rather have to save the policy to dictate when I get to do the comparison.''} [E4]. They also viewed notes summarizing policy and response changes as potentially unnecessary to reduce clutter in the sidebar. We incorporated this feedback into the final design of our system. The facilitator's guide for this co-design session can be found in the ``Co-design workshop 3'' section of our Supplementary Materials.

\section{Connections Between UX and LLM Policy Prototyping}
See Table \ref{t:mapping}.

\label{a:ux-polipro}

\begin{table*}[h]
\centering
\small
\label{tab:ux-policy-translation}
\begin{tabular}{p{3cm}p{2.3cm}p{3.5cm}p{3.5cm}p{3cm}}
\toprule
\textbf{Observation} & \textbf{Relevant UX Method} & \textbf{UX Definition \& Usage} & \textbf{Usage in LLM Policy Prototyping} & \textbf{\pp Features} \\
\midrule

\textbf{Incomplete feedback loops} (Section \ref{form:prototyping}) & 
\textbf{Rapid Prototyping} (e.g., \cite{lim2008anatomy, houde1997prototypes, buchenau2000experience, camburn2017design, dow2010parallel, quicksey-policy-prototypes, Kontschieder-policy-prototyping}) & 
Tight feedback loops of \textbf{ideating, implementing, and evaluating design ideas}. Allows designers to identify usability issues early, explore alternatives, and align teams to shared visions & 

Tight feedback loops of \textbf{ideating, drafting, and testing policy statements} for quick identification of policy ``usability'' issues (e.g., unclear statements), characteristics of responsible model behavior, and translations of that behavior into policy. & 

Response (re)generation with policy-informed model, policy suggestion upon editing response. \\

\midrule

\textbf{Operating at both high and low levels} (Section \ref{form:heuristics}) & 
\textbf{Low-fidelity prototyping} (e.g., \cite{rudd1996low, virzi1996usability, walker2002high}) & 
Artifact \textbf{loosely} resembling the final product in terms of look \& feel and/or implementation. Cheap to create and iterate upon, ideal for collecting early requirements and feedback. & 

Artifact providing \textbf{high-level documentation} of responsible model behavior to quickly gather and integrate perspectives/feedback. Requires focus on \textbf{high-level} details. \textbf{We focus on this type of policy prototype in our work.} & 

Heuristics editor and checker, freeform document editor. \\

\midrule

\textbf{Operating at both high and low levels} (Section \ref{form:heuristics}) & 
\textbf{High-fidelity prototyping} (e.g., \cite{rudd1996low, virzi1996usability, walker2002high}) & 
Artifact \textbf{closely} resembling the final product in terms of look \& feel and/or implementation. They are useful for collecting detailed feedback but may be expensive to create.  & 

Artifact providing \textbf{detailed documentation} of responsible model behavior to guide alignment efforts, often with polished wording, illustrative examples, legal sign-off, and more. \textbf{Requires focus on high- and low-level details.} Example: OpenAI Model Spec \cite{oai-model-spec}. & 

N/A---focus of \pp is low-fidelity prototyping. \\

\midrule

\textbf{Scenarios grounded discussions} (Section \ref{form:scenarios}) & 
\textbf{Storyboarding/ scenario-building} (e.g., \cite{figma-storyboards, truong2006storyboarding, hooper1982scenario, bodker1999scenarios, andersen1999scenario}) & 
Concrete representations of \textbf{users, contexts, and tasks} to ground abstract design ideas. \textbf{Panels} add context and illustrate user stories. Promotes reflection and communication among stakeholders. & 

\textbf{Sample user-AI conversations} to ground policy discussions and creation. \textbf{Conversational turns} add context and illustrate sample user \& model behaviors. Promotes reflection and communication among stakeholders. &
Interactive scenarios, adding/extending scenarios, spotlight scenarios. \\

\midrule

\textbf{Experts valued synchronous collaboration} (Section \ref{form:synchronous}) & 
\textbf{Design workshops} e.g., \cite{rosner2016out, elsden2020design, rosenbaum2002focus, kimbell2017prototyping, hagan2021prototyping} & 
Common collaborative method for gathering user requirements, studying empirical phenomena, and evaluating interactive systems. Can serve as a field site, research instrument, or a research account. & 

Small-group sessions that serve as a \textbf{field site} for collective ideation and reflection of responsible model behavior in domain-specific use cases. &
Collaborative multi-player editor, spotlight scenarios, response flagging. \\

\bottomrule
\end{tabular}
\caption{Mapping of UX methods relevant to insights from our observational study (Section \ref{s:formative}) to their usage in LLM policy prototyping, to features in \pp supporting that usage.}
\Description{A 5-column table with 5 rows, mapping UX concepts of rapid prototyping, low-fidelity prototyping, high-fidelity prototyping, storyboarding, and design workshops to their usage in LLM policy prototyping.}
\label{t:mapping}
\end{table*}

\section{Starter Heuristics and Policy Objectives Section}
\label{a:starter-material}
Heuristics:
\begin{enumerate}
    \item Policy statements should be written clearly and precisely.
    \item If a policy statement applies in some scenarios but not others, its scope should be communicated clearly.
    \item The policy should incorporate insights from real-world professional practices to guide appropriate and responsible behavior.
\end{enumerate}

\noindent Objectives:
\begin{itemize}
    \item Help users achieve their goals (if applicable) by following instructions and providing helpful responses.
    \item Consider potential benefits and harms to a broad range of stakeholders.
    \item Respect social norms and applicable law.
\end{itemize}

\section{Post-Study Policy Rating Questions}
\label{a:policy-likert-qs}
All questions were on a 5-point Likert scale.
\begin{itemize}
    \item Please rate the extent you think this policy addresses important considerations of AI behavior within your professional domain.
    \item Please rate the extent to which you agree with this policy. By agreement, we mean whether you can see yourself taking (or aspire to take) a similar approach if you were drafting the same policy.
    \item Here are some heuristics the policy was supposed to satisfy. 1) Policy statements should be written clearly and precisely. 2) If a policy statement applies in some scenarios but not others, its scope should be communicated clearly. 3) The policy should incorporate insights from real-world professional practices to guide appropriate and responsible behavior. Do you think the policy did a good job at satisfying these heuristics?
\end{itemize}

Note that we did not analyze and report on the third question because it became apparent during the study that not all experts agreed with these heuristics. Thus, a high rating on this question might not have as positive of a signal as we assumed it would. 

\section{Evaluation Study Participants}
See Table \ref{t:eval-participants}.
\label{a:eval-participants}

\begin{table*}[h]
\centering
 \begin{tabular}{p{0.5cm} p{0.5cm} p{1.5cm} p{1.5cm} p{1cm} p{5cm} p{2cm}}
 \hline
 \bfseries G\# & \bfseries P\#& \bfseries Gender & \bfseries Age Range & \bfseries YoE & \bfseries Education Status & \bfseries GenAI Use\\ 
 \hline
 \multirow{3}{=}{MH1} & P1 & Man & 25--34 & 3 & Clinical Psychology Ph.D. (in-progress) & Regular \\ 
 & P2 & Man & 35--44 & 15 & Clinical Psychology Psy.D. (completed) & Regular \\
 & P3 & Man & 25--34 & 4 & Clinical Psychology Ph.D. (in-progress) & Regular \\
 \hline
 \multirow{2}{=}{MH2} & P4 & Woman & 25--34 & 7 & Clinical Psychology Ph.D. (in-progress) & Regular \\ 
 & P5 & Woman & 25--34 & 4 & Clinical Psychology Ph.D. (in-progress) & Occasional \\
 \hline
 \multirow{3}{=}{MH3} & P6 & Woman & 35--44 & 10 & Clinical Psychology Master's (completed) & Occasional \\ 
 & P7 & Woman & 45--54 & 15 & Clinical Psychology Psy.D. (completed) & Regular \\
 & P8 & Woman & 45--54 & 25 & Clinical Psychology Psy.D. (completed) & Regular \\
 \hline
 \multirow{2}{=}{MH4} & P9 & Woman & 25--34 & 8 & Clinical Psychology Ph.D. (in-progress) & Occasional \\ 
 & P10 & Woman & 45--54 & 14 & Clinical Psychology Ph.D. (completed) & Regular \\
 \hline
 \multirow{3}{=}{L1} & P11 & Man & 25--34 & 3 & J.D. (completed) & Regular \\ 
 & P12 & Woman & 18--24 & 4 & LL.M. (in-progress) & Regular \\
 & P13 & Woman & 25--34 & 10 & Law Ph.D. (in-progress) & Regular \\
 \hline
 \multirow{3}{=}{L2} & P14 & Man & 25--34 & 2 & LL.M. (in-progress) & Regular \\ 
 & P15 & Woman & 18--24 & 3 & LL.M. (in-progress) & Regular \\
 & P16 & Woman & 25--34 & 10 & LL.M. (in-progress) & Regular \\
 & P17 & Man & 25--34 & 5 & LL.M. (in-progress) & Regular \\
 \hline
 \multirow{3}{=}{L3} & P18 & Man & 25--34 & 13 & Law Master's (completed) & Regular \\ 
 & P19 & Woman & 25--34 & 4 & LL.M. (in-progress) & Regular \\
 & P20 & Woman & 25--34 & 3 & LL.M. (completed) & Regular \\
 \hline
 \multirow{3}{=}{L4} & P21 & Man & 25--34 & 4 & J.D. (in-progress) & Regular \\ 
 & P22 & Man & 18--24 & 6 & LL.M. (in-progress) & Occasional \\
 \bottomrule
\end{tabular}
\caption{Details of participants (gender, age range, education status, years of practical experience, and generative AI use) in our evaluation study. The ``GenAI use'' column refers to participants' experience using generative AI tools, whether it be personally or professionally, as determine by their frequency of use. The response ``Occasional'' corresponds to the following description: ``I've tried it here and there but don't use it regularly.'' All participants specializing in mental health were based in the US. All participants specializing law except two (who were based in Europe and Asia, respectively) were based in the US.}
\Description{A table summarizing the background information of participants (P1 to P22) grouped into eight groups in our evaluation study.}
\label{t:eval-participants}
\end{table*}

\section{Areas of Disagreement Across Expert Groups}
\label{a:disagreement}
Within \textbf{mental health groups}, there was some disagreement over 1) whether the model should act like a therapist, and 2) the appropriate conversational tone before a proper assessment of the user is made. Experts in some groups debated over the question from 1) and concluded that a model acting like a therapist may be a temporary solution until they can access professional support, which they recognized can come with long waits. For 2), some suggested that the model should keep responses generic until a proper assessment of the user could be made, while others believed that the model's level of empathy should depend on the user's level of expressed distress. 

Within \textbf{legal groups}, experts disagreed over whether the model should suggest action items for the user. Some considered it irresponsible for the model to make any conclusions about potential legal actions to pursue, while others acknowledged that AI systems legally bound in the same way lawyers are and therefore did not take issue with AI-recommended actions. The latter contrasts with findings from Cheong et al. \cite{cheong2024not}, where legal experts unanimously agreed that AI should not recommend actions. 

\textbf{Across the domains}, disagreements arose over whether the model should, under any circumstances, attempt to mimic a human professional. While cases for and against were made among mental health experts, there was broad consensus across legal experts that the model should not act like a lawyer. Further, the level of empathy expressed by the model was another point of disagreement---empathetic responses were seen as essential in mental health and undesirable in legal settings.

Overall, we expect that some of these disagreements may be resolved with further iterative prototyping of policies. Once some areas of disagreement have been isolated, further rounds of policy prototyping can be conducted using scenarios \textit{specifically crafted to target these disagreements}. For example, while mental health experts did not initially agree on whether the model should act like a therapist, policy prototyping with more scenarios featuring a therapist-like model may actually reveal significant agreement about specific circumstances under which the model should exhibit that behavior. While we did not have time for more sessions with our groups, we see promise in using multiple rounds of policy prototyping with carefully chosen scenarios to shed more light on strategies for resolving these disagreements.

\end{document}